\documentclass[preprint,twocolumn,linenumbers=false]{aastex631}
\usepackage{bm}
\usepackage{threeparttable}
\usepackage{amsmath}
\usepackage{graphicx}
\hypersetup{
  colorlinks=true,
  urlcolor=blue,
}

\begin{document}

\title{Atmospheric Circulation of High-Obliquity Mini-Neptunes}

\author[0000-0001-9700-9121]{Yanhong Lai}
\affiliation{Tsung-Dao Lee Institute, Shanghai Jiao Tong University, 1 Lisuo Road, Shanghai 200127, People’s Republic of China}
\email{yanhonglai@sjtu.edu.cn}

\author[0000-0003-2278-6932]{Xianyu Tan}
\affiliation{Tsung-Dao Lee Institute, Shanghai Jiao Tong University, 1 Lisuo Road, Shanghai 200127, People’s Republic of China}
\affiliation{School of Physics and Astronomy, Shanghai Jiao Tong University, 800 Dongchuan Road, Shanghai 200240, People’s Republic of China}
\affiliation{State Key Laboratory of Dark Matter Physics, Shanghai Jiao Tong University, 1 Lisuo Road, Shanghai 200127, People’s Republic of China}
\email{xianyut@sjtu.edu.cn}

\author[0000-0001-8283-3425]{Yubo Su}
\affiliation{Department of Astrophysical Sciences, Princeton University, 4 Ivy Ln, Princeton, NJ 08544, USA}
\affiliation{Canadian Institute for Theoretical Astrophysics, 60 St. George Street, Toronto, ON M5S 3H8, Canada}

\begin{abstract}
With the operation of JWST, atmospheric characterization has now extended to low-mass exoplanets. In compact multiplanetary systems, secular spin-orbital resonance may preserve high obliquities and asynchronous rotation even for tidally-despinning, low-mass planets, potentially leading to unique atmospheric circulation patterns. 
To understand the impact on the atmospheric circulation and to identify the potential atmospheric observational signatures of such high-obliquity planets, we simulate the three-dimensional circulation of a representative mini-Neptune K2-290 b, whose obliquity may reach about 67$^{\circ}$. 
Whether synchronously rotating or not, the planet’s slow rotation, moderate temperature and radius result in a global Weak-Temperature-Gradient (WTG) behavior with moderate horizontal temperature contrasts. Under synchronous rotation, broad eastward superrotating jets efficiently redistribute heat. Circulation in an asynchronous rotation exhibits a seasonal cycle driven by high obliquity, along with quasi-periodic oscillations in winds and temperatures with a period of $\sim$70 orbital periods. These oscillations, driven by wave-mean flow interactions, extend from low- to mid-latitudes due to the slow planetary rotation. 
Higher atmospheric metallicity strengthens radiative forcing, increasing temperature contrasts and jet speeds. Clouds have minimal impact under synchronous rotation but weaken jets under nonsynchronous rotation by reducing temperature contrasts. 
In all cases, both thermal emission and transmission spectra exhibit moderate observational signals at a level of 100 ppm, and high-obliquity effects contribute differences at the $\sim$10 ppm level. Our results are also applicable to a range of potential high-obliquity exoplanets, which reside in the WTG regime and likely exhibit nearly homogeneous horizontal temperature patterns.  
\end{abstract}


\section{Introduction}
\label{sec:intro}

Since the discovery of the first exoplanet around sun-like stars by \cite{Mayor1995}, over 6000 exoplanets have been confirmed to date\footnote{\url{https://exoplanetarchive.ipac.caltech.edu}}. Among those exoplanets, mini-Neptunes with radii between 2 and 4 Earth radii are the most common type \citep{fressin2013,fulton2017}. With the operation of JWST, in-depth characterization of mini-Neptunes becomes accessible, which opens up a new window for probing the formation, long-term evolution, atmospheric and interior processes of these worlds \citep{kempton2024}. 

Phase-curve observations provide one of the most comprehensive methods for characterizing the atmospheres of close-in exoplanets. In addition, the eclipse mapping technique emerges as a new tool to probe the detailed two-dimensional dayside patterns of these planets by analyzing the time-dependent flux variations during a planet’s ingress and egress of secondary eclipse \citep{williams2006,rauscher2007,majeau2012,coulombe2023broadband,challener2024,challener2025}. Interpreting the phase curve and eclipse observations often relies on modeling the three-dimensional (3D) atmospheric structure and winds of these planets using general circulation models (GCMs).  Most existing GCM studies assume that close-in mini-Neptunes rotate synchronously, with the planetary spin axis the same as the planetary orbital axis \citep[e.g.][]{innes2022,landgren2023}. 
In a synchronously rotating state, the circulation patterns of mini-Neptunes share similar properties to those of synchronously rotating hot/warm Jupiters, which have been extensively explored over two decades \citep{showman2002,showman2011,pierrehumbert2019,showman2020review}. Classically, planetary-scale, standing Rossby and Kelvin waves, driven by a stationary day-night temperature contrast, transport eddy momentum toward the equator, resulting in an eastward superrotating jet around the equator. 

The synchronous rotation of close-in exoplanets is generally motivated by arguing that the tidal spin synchronization timescale is much shorter than the age of the system \citep{guillot1996}.
However, more sophisticated theoretical analyses suggest that this argument may be insufficient in multiplanetary systems, where a variety of resonances can prevent the planet from reaching the synchronized state.
For instance, one class of resonances between the secular precessions of a planet's spin and orbital axes can trap a planet's spin in a high-obliquity (i.e.\ a large misalignment between the planet's spin axis and its orbit normal), subsynchronous spin state \citep[e.g.][]{winn2005, fabrycky2007, millholland2018, su2022a, su2022b}.
In this mechanism, only a small mutual inclination of the planets' orbits can result in substantial asynchronicity.
Numerous other pathways to avoid synchronization involve more complex physical phenomena such as large eccentricities \citep[e.g.][]{valente2022},
complex tidal and rheological models \citep[e.g.][]{correia2001_venus, boue2016_maxwell, revol2023_venus},
and rigid-body dynamics \citep[e.g.][]{goldreichpeale1966, correialaskar2004_mercury, yuan2024}.
While measurements of exoplanet obliquity and rotation rate are difficult, the first constraint on the obliquities of super-Jupiter-mass objects have recently become possible by combining three key observables: projected rotational velocities $v\sin{i}$ from the broadening of spectral lines, rotation periods from photometry, and orbital inclinations \citep{bryan2020obliquity, bryan1987parameter, palmabifani2023_obl, poon2024_obl}.

Planetary obliquity could exert a significant influence on atmospheric circulation, exoplanet habitability, and observational signatures, which has been widely explored on habitable-zone terrestrial planets \citep[e.g.][]{dobrovolskis2009,ferreira2014,linsenmeier2015,wang2016,kilic2017,kang2019,kodama2022,jernigan2023} and warm Jupiters \citep{rauscher2017,ohno2019a,ohno2019b,adams2019,albrecht2022,rauscher2023}. Obliquity affects the meridional distribution of stellar insolation, resulting in seasonal variation of atmospheric circulation. 
In particular, the polar regions could receive more stellar energy than the equatorial regions when obliquity exceeds 54$^{\circ}$, which leads to a reversed distribution of stellar isolation compared to that with low obliquities \citep{ferreira2014,linsenmeier2015}. Under this circumstance, the meridional temperature gradient may reverse, causing the dominant atmospheric winds to shift from eastward to westward \citep{rauscher2017,ohno2019a}, as expected from thermal wind balance that relates meridional temperature contrasts to the vertical shear of the zonal wind \citep{vallis2017atmospheric}.
Furthermore, based on shallow-water models and with a focus on the fast-rotating warm Jupiters, \cite{ohno2019a} suggested that the atmospheric circulation of exoplanets with non-zero obliquities could be categorized into five different dynamic regimes, dependent on the radiative timescale, orbital period, planetary rotation period, and obliquity. 

Previous studies mentioned above mainly focused on the atmospheric circulation of high-obliquity, fast-rotating hot/warm Jupiters and terrestrial planets with obliquities and rotation rates arbitrarily chosen. In this work, we are interested in the atmospheric circulation of high-obliquity slow-rotating mini-Neptunes whose potential asynchronicity is well motivated and calculated by theories of secular spin-orbit resonance \citep{su2022a,su2022b}. We also hope to identify potential atmospheric observational signatures of such high-obliquity mini-Neptunes, by employing the GCM,  ADAM (ADvanced Atmospheric MITgcm, formerly known as SPARC/MITgcm), to simulate the 3D atmospheric circulation of a representative mini-Neptune K2-290 b, whose obliquity could reach about 67$^{\circ}$ and spin frequency could be about 0.69 of the orbital frequency (values from Equation~\eqref{eq:tce2} in Section \ref{sec:cassini_states}). The organization of this paper is as follows. We introduce the secular spin-orbit theory, atmospheric model setup and experimental design in Section \ref{sec:model}. We present the main results of synchronously and nonsynchronously rotating cases in Section \ref{sec:sync} and \ref{sec:nonsync}. The influence of higher metallicity and clouds are presented in Section \ref{sec:10solar_cloud}. We discuss the observational signatures and model limitations in Section \ref{sec:discuss} and summarize in Section \ref{sec:conclude}.

\section{Model Setup and Experimental design}\label{sec:model}

\subsection{Theory of high obliquity of close-in exoplanets}\label{sec:cassini_states}

In this section, we briefly outline the key physical effects that can lead to a close-in exoplanet retaining a large obliquity and a subsynchronous rotation rate in spite of ongoing tidal dissipation.
Our discussion largely summarizes the important results from \citet{su2022a}, and we restrict our attention to two-planet systems for simplicity and concreteness (though the processes described here are also important in systems with three or more planets, see e.g.\ \citealp{saillenfest2019_seculardyn} and \citealp{su2022b}).

Consider two planets with masses $m$ (the inner planet, whose spin evolution is studied in this work) and $m_{\rm p}$ (the outer, perturbing planet) in orbit around its host star with mass $M_\star$ on circular orbits with semi-major axes $a$ and $a_{\rm p}$.
The mutual inclination between the orbits is denoted $I$.
We focus our attention on the spin dynamics of the inner planet, and we denote its radius by $R$, its obliquity $\phi$, and its spin-orbit frequency ratio $\Omega_n \equiv \Omega / n$ where $\Omega$ is the spin frequency of the planet and $n$ its orbital frequency.
The spin dynamics of the inner planet depend on two precessional effects: (i) the precession of the planet's spin axis, driven by the gravity of the host star on the planet's rotationally-induced quadrupole moment $J_2 \equiv (k_2/3) \Omega^2 R^3 / Gm$ where $k_2$ is the planet's second Love number; and (ii) the precession of the planet's orbital plane, driven by the misaligned orbit of the outer companion.
The combination of these two precession effects describes the system known as Colombo's Top \citep[e.g.][]{colombo1966, peale1969, ward1975tidal}, whose equilibria are the well-known Cassini States.

In addition to these effects, tidal dissipation in the planet induced by the host star will act to drive the planet's spin frequency towards its orbital frequency and to drive its obliquity towards zero.
As shown in \citet{su2022a}, the inclusion of tidal dissipation in the Colombo's Top system reduces the number of long-lived equilibrium spin states for the planet to no more than two.
The number and properties of these equilibria depends on the parameter
\begin{equation}
    \eta_{\rm sync}
        =
            f(\alpha) \frac{J}{L_{\rm p}}
            \frac{3C}{2k_2}
            \frac{mm_{\rm p}}{M_\star^2}
            \left(\frac{a}{a_{\rm p}}\right)^3
            \left(\frac{a}{R}\right)^3
            \cos I
            ,
\end{equation}
where $f(\alpha) \equiv b_{3/2}^{(1)}(\alpha) / (3\alpha) \approx 1 + 15\alpha^2/8$ is related to the Laplace coefficient $b_{3/2}^{(1)}$, $\alpha = a / a_{\rm p}$, $L_{\rm p} \approx m_{\rm p}\sqrt{GM_\star a_{\rm p}}$ is the angular momentum of the outer companion, $J$ is the total angular momentum of the two planets, and $C = \mathcal{I} / mR^2$ is the normalized moment of inertia of the planet, where $\mathcal{I}$ is the planet's moment of inertia (we have adopted the expressions from Section~5.3 of \cite{su2022a}, which are most appropriate for the compact hierarchies considered in this work).
$\eta_{\rm sync}$ has the physical interpretation of being the ratio of the planet's spin precession frequency to that of its orbit when the planet's spin is synchronized ($\Omega_n = 1$).
When $\eta_{\rm sync} \gtrsim 1$, the only planetary spin equilibrium has a low obliquity\footnote{
$\approx I$, up to corrections of order $\sim \eta_{\rm sync}^{-1}$ \citep{su2020_disk}.}
and rotates nearly synchronously, but a high-obliquity equilibrium spin state appears when $\eta_{\rm sync} \lesssim 1$.
The obliquity $\phi$ and spin-orbit frequency ratio $\Omega_n$ in this equilibrium can be solved exactly by numerical identification of the Cassini State equilibria, though useful approximate forms in the limit $\eta_{\rm sync} \ll 1$ are given by \citep{su2022a}
\begin{align}
    \cos\phi &\approx \sqrt{\frac{\eta_{\rm sync} \cos I}{2}},&
    \Omega_n &\approx \sqrt{2 \eta_{\rm sync} \cos I}.\label{eq:tce2}
\end{align}
It can be seen that, when $\eta_{\rm sync} \lesssim 1$, the planet's obliquity is large and its spin rate is slow.

However, when tidal dissipation is too strong, it can render the high-obliquity state unstable \citep{levrard2007, fabrycky2007}.
This condition can be succinctly expressed as requiring that the planet's tidal evolution timescale $t_{\rm s}$ be longer than some critical timescale $t_{\rm s, c}$, where \citep{su2022a, guerrero2023}
\begin{align}
    \frac{1}{t_{\rm s}}
        &=
            \frac{3k_2}{4CQ}
            \frac{M_\star}{m}
            \left(\frac{R}{a}\right)^3
            n,\\
    \frac{1}{t_{\rm s, c}}
        &=
            \frac{3m_{\rm p}}{4M_\star}
            \left(\frac{a}{a_{\rm p}}\right)^3
            n \sin I
            \sqrt{\frac{\eta_{\rm sync}\cos I}{2}}
            ,\label{eq:tidal_break}
\end{align}
where $Q$ is the planet's tidal dissipation quality factor.
If $t_{\rm s} < t_{\rm s, c}$, then the tidal obliquity damping overpowers the obliquity excitation of the resonant Cassini State, and no stable high-obliquity state exists.

\begin{table*}[!t]
\caption{
Physical parameters for high-obliquity mini-Neptune candidates that are in both the NASA Exoplanet Archive \citep{nasa_exoplanetarchive} and TEPCat \citep{tepcat} as two-planet systems.
\label{tab:list}}
\centering
\begin{tabular}{lccccccccc}
\hline
Planet & d [pc] & T$_{eq}$ [K] & R [R$_E$] & g [g$_E$] & a [AU] & $\Omega_n$ & P$_{\rm orb}$ [days] & $\phi$ [$^\circ$] & $\wedge$\\
\hline
Kepler-241 b & 512  & 559  & 2.33 & 1.55 & 0.0941 & 0.866 & 12.7 & 55 & 17\\
Kepler-263 b & 755  & 584  & 2.66 & 1.44 & 0.1202 & 0.906 & 16.6 & 51 & 20\\
Kepler-258 b & 576  & 625  & 4.06 & 1.17 & 0.1032 & 0.867 & 15.5 & 55 & 13\\
Kepler-261 b & 317  & 679  & 2.17 & 1.60 & 0.0878 & 0.810 & 10.4 & 59 & 17\\
Kepler-284 b & 1032 & 696  & 2.24 & 1.58 & 0.1035 & 0.761 & 12.7 & 62 & 20\\
Kepler-209 b & 577  & 704  & 2.26 & 1.57 & 0.1227 & 0.995 & 16.1 & 25 & 25\\
Kepler-242 b & 600  & 732  & 2.61 & 1.45 & 0.0744 & 0.951 & 8.2  & 43 & 11\\
Kepler-1129 b & 1838 & 737  & 5.08 & 1.04 & 0.1644 & 0.938 & 24.3 & 46 & 17\\
Kepler-165 b  & 561  & 742  & 2.32 & 1.55 & 0.0722 & 0.952 & 8.2  & 43 & 13\\
Kepler-1093 b & 1073 & 776  & 2.26 & 1.57 & 0.1747 & 0.893 & 25.1 & 52 & 41\\
Kepler-175 b  & 1312 & 822  & 2.56 & 1.48 & 0.1053 & 0.709 & 11.9  & 65 & 18\\
Kepler-161 b  & 434 & 839  & 2.12 & 1.62 & 0.0537 & 0.985 & 4.92  & 33 & 9\\
Kepler-182 b  & 1549 & 937  & 2.58 & 1.47 & 0.0955 & 0.886 & 9.8  & 53 & 15\\
K2-243 b     & 267  & 1035 & 2.17 & 1.60 & 0.1087 & 0.838 & 11.5 & 57 & 23\\
Kepler-1642 c & 414  & 1095 & 4.81 & 1.07 & 0.0663 & 0.858 & 6.7  & 56 & 6\\
\textbf{K2-290 b} & \textbf{273}  & \textbf{1230} & \textbf{3.06} & \textbf{1.34} & \textbf{0.0923} & \textbf{0.685} & \textbf{9.2} & \textbf{67} & \textbf{14}\\
\hline
\end{tabular}
\begin{tablenotes}
\footnotesize
\item  Note: $d$ is the distance from Earth, $T_{eq}$ is the global radiative equilibrium temperature with zero albedo, $R$ and $R_E$ are radii of planet and Earth, $g$ and $g_E$ are gravity of planet and Earth, $a$ is semi-major axis, $\Omega_n$ is spin/orbit frequency ratio, $P_{\rm orb}$ is orbital period, $\phi$ is planetary obliquity, and $\wedge$ is the Weak-Temperature-Gradient (WTG) parameter (Equation~(\ref{eq:WTGpara})).
The observed properties are taken from TEPCat, and $\phi$ and $\Omega_n$ are evaluated from Equation~\eqref{eq:tce2}.
K2-290 b is highlighted as our model target in this study for its close distance to Earth and high equilibrium temperature, which are favored by atmospheric observations.
The parameters for K2-290b are taken from \citet{hjorth2021_k2290}.
\end{tablenotes}
\end{table*}

To identify systems that could host high-obliquity mini-Neptunes due to the processes described above, we collected all two-planet systems in the NASA Exoplanet Archive \citep{nasa_exoplanetarchive} as of June 25, 2025, numbering $603$ in total.
We then selected for systems where the inner planet satisfies $t_{\rm s} > t_{\rm s, c}$; fiducial values of $Q = 10^3$, $C = 0.3$, and $k_2=0.4$ are adopted, as these values lie roughly between those of Earth and the solar system ice giants \citep{lainey2016_tidal, nimmo2023_uranusQ}, while $I = 2^\circ$ is taken from \citet{zhu2018_multiplicity_inc}.
For systems with multiple observational constraints on the system parameters, we take the average of the reported values weighted by the reciprocals of squared uncertainties \citep[e.g.,][]{taylor_erroranalysis}.
We then select for systems with $\eta_{\rm sync} \lesssim 1$ (potentially high-obliquity systems) and with $R > 2R_\oplus$ (inner mini-Neptunes), resulting in a total of $36$ potentially high-obliquity mini-Neptunes.
We repeat the same procedure using the TEPCat catalog \citep{tepcat}: a total of $16$ systems are identified as candidate high-obliquity two-planet systems in both catalogs\footnote{Most rejected systems have discrepant multiplicities between the two catalogs due to newer and/or lower-confidence detections included in the Exoplanet Archive.}.
We report the respective obliquities and spin frequencies of these systems in ascending order of their equilibrium temperatures $T_{\rm eq}$ in Table~\ref{tab:list}.
Note that while different choices of $C$, $k_2$, and $I$ may change the specific obliquities predicted for a given system, the typical order of magnitude of the spin rates and obliquities of high-obliquity systems changes little and in predictable ways following Equation~\eqref{eq:tce2}.
Adopting a lower (or higher) value for $Q$ will make a moderately more significant impact by decreasing (or increasing) the maximum tidally-stable obliquity and subsynchronicity.
Finally, note that the uncertainty on the estimated $\Omega_n$ and $\cos \phi$ of K2-290b are estimated to be $\sim 20\%$, dominated by the uncertainty on the mass (via the radius) of K2-290b itself.
Many of the other systems have much larger fractional uncertainties than K2-290 due to poorly constrained masses.

\subsection{The ADAM General Circulation Model}\label{sec:sparc}
Following \cite{mehta2025}, we introduce the term ADAM (ADvanced Atmospheric MITgcm)\footnote{This honours the late Adam Showman, whose pioneering contributions laid much of the conceptual and practical groundwork for exoplanet atmospheric dynamics.} as an umbrella designation for exoplanet modelling frameworks built upon the MITgcm \citep{adcroft2004} and the non-grey radiative transfer code of \cite{marley1999}. Under this naming convention, model configurations are referenced by specifying the relevant physical modules, e.g., ADAM with SPARC, ADAM with double-grey radiative, ADAM with double-grey and H$_2$-dissociation, or ADAM with SPARC, active tracer clouds, and customized opacities, and so on. This nomenclature provides a unified and transparent framework for describing the growing suite of MITgcm-based exoplanet modelling capabilities. To avoid ambiguity with other MITgcm-based exoplanet modelling efforts, we clarify that ADAM refers specifically to the lineage of models descending from the developments initiated by \cite{showman2009} and extended through a series of key advancements afterwards \citep[e.g.,][]{lewis2010,kataria2013,parmentier2013,komacek2017,tan2019uhj,tan2021a,komacek2022cloud,steinrueck2023,mehta2025}. More details are referred to \cite{mehta2025}. 

We model the atmospheric circulation of mini-Neptunes using ADAM with SPARC, active tracer clouds. The dynamical core solves the hydrostatic primitive equations on a cubed sphere grid in pressure coordinates \citep{adcroft2004}:
\begin{equation}
    \frac{d\textbf{v}}{dt}=-f\hat{k}\times\textbf{v}-\nabla_p \Phi + R_v, 
    \label{eq1}
\end{equation}
\begin{equation}
    \frac{\partial \Phi}{\partial p} = -\frac{1}{\rho},
    \label{eq2}
\end{equation}
\begin{equation}
    \nabla_p \cdot \textbf{v} + \frac{\partial w}{\partial p} =0,
    \label{eq3}
\end{equation}
\begin{equation}
    \frac{d\theta}{dt}= \frac{g\theta}{c_p T}\frac{\partial F}{\partial p} + R_{\theta}, 
    \label{eq4}
\end{equation}
where $\textbf{v}$ is the horizontal velocity, $\omega=dp/dt$ is the vertical velocity in pressure coordinates, $p$ is pressure, $f=2\Omega\sin\phi$ is the Coriolis parameter, $\Omega$ is the rotation rate, $\phi$ is latitude, $\Phi$ is geopotential, $\rho = p/ RT$ is gas density following the ideal gas law, $R$ is the specific gas constant, $T$ is temperature, $\theta=T\left(p_s/p\right)^{R/c_p}$ is potential temperature, $p_s$ is a reference pressure, $c_p$ is heat capacity at constant pressure, $F$ is net radiative flux including the incoming stellar flux and atmospheric thermal flux, $g$ is gravity, $\frac{d}{dt}=\frac{\partial}{\partial t} + \textbf{v}\cdot \nabla_p + \omega \frac{\partial}{\partial p}$ is the total derivative, and $\nabla_p$ is the horizontal gradient over constant pressure levels.

The net radiative flux $F$ in each column is calculated using the non-grey radiative transfer model of \cite{marley1999}, which solves the two-stream radiative transfer equations based on the correlated-$k$ method. We assumed equilibrium chemistry in the radiative transfer calculation as in \cite{showman2009}. $R_v$ is a Rayleigh drag employed in the deep layers where pressure exceeds 100 bars to reduce the horizontal winds to zero within a prescribed drag timescale of 10$^5$ s, which roughly represents the effect of momentum mixing between the deep interior and the weather layer. $R_{\theta}$ represents the thermal heating converted from kinetic energy by the Rayleigh friction.

\subsection{Cloud formation scheme}\label{sec:cloudscheme}
For simulations with clouds included, the cloud formation scheme is similar to that of \cite{tan2021a,tan2021b}. Two tracer equations representing the mixing ratio of the condensable vapor ($q_v$, mass ratio between condensable vapor to the background dry gas) and the mixing ratio of cloud particles ($q_c$) for one species are integrated along with the primitive equations:
\begin{equation}
    \frac{dq_v}{dt}=-\delta \frac{q_v - q_c}{\tau_c} + (1-\delta)\frac{\min(q_s-q_v, q_c)}{\tau_c} + Q_{deep}, 
    \label{eq5}
\end{equation}
\begin{equation}
    \frac{dq_c}{dt}=\delta \frac{q_v - q_c}{\tau_c} - (1-\delta)\frac{\min(q_s-q_v, q_c)}{\tau_c} -\frac{\partial (\langle q_c  V_s\rangle)}{\partial p}, 
    \label{eq6}
\end{equation}
where $\tau_c = 100$ s is the conversion timescale between cloud and condensable vapor, $q_s$ is the saturated mixing ratio, $V_s$ is the settling speed of cloud particles (for details, refer to \cite{parmentier2013}) and $\langle q_c V_s \rangle$ is normalization over the particle size distribution, $Q_{\rm deep} = -(q_v-q_{\rm deep})/\tau_{\rm deep}$ is the vapor source applied only in pressure regions deeper than 100 bars, which relaxes the vapor mixing ratio toward the deep mixing ratio $q_{\rm deep}$ within a timescale $\tau_{\rm deep}$ of 1000 s. 

The first and second terms on the right-hand side of Equations (\ref{eq5})-(\ref{eq6}) represent the sources/sinks due to condensation and evaporation, respectively. When $q_v$ is less than $q_s$, evaporation occurs and $\delta$ is set to 0. Otherwise, when $q_v$ exceeds $q_s$, cloud forms and $\delta$ is set to 1. 
The saturation mixing ratio $q_s =P_T q_{\rm deep}/p $, where $P_T$ is the total gas pressure (in unit of bars) at which tracer
saturates, dependent on temperature and cloud composition \citep{visscher2010,morley2012}. In this study, we include two cloud species that may be relevant at this temperature: MnS and Na$_2$S. Given the lack of atmospheric measurements to constrain the metallicity of K2-290 b, the cloud-included models are computed assuming solar metallicity, following common GCM practices \citep{showman2020review}. Based on solar elemental ratios \citep{lodders2003} and assuming that all Na and Mn condense into these two species, the $q_{\rm deep}$ is set to be 2.17$\times10^{-5}$ kg\,kg$^{-1}$ for MnS and 5.9$\times10^{-5}$ kg\,kg$^{-1}$ for Na$_2$S. 
It is noteworthy that $q_{\rm deep}$ will linearly increase with higher metallicty, supplying more deep-layer vapor for cloud formation, which introduce stronger cloud opacity and affect the condensation level of clouds, both of which are expected to alter the thermal structure and circulation \citep{ohno2018}.

\begin{table*}[!t]
\caption{Summary of simulations and model parameters\label{tab:paras}}
\centering
\begin{tabular}{llll}
\hline\hline
\multicolumn{4}{l}{\textbf{Basic planetary parameters}} \\
\hline
Parameter & Value & Cases \\
\hline
Obliquity & 0$^{\circ}$, 67$^{\circ}$ & Synchronous / nonsynchronous \\
Orbital period & 9.2 days &  \\
Rotation period & 9.2, 13.4 days & Synchronous / nonsynchronous \\
\hline
\multicolumn{4}{l}{\textbf{1. Control}} \\
\hline
Heat capacity & 13000 J\,kg$^{-1}$\,K$^{-1}$ & Specific gas constant & 3714 J\,kg$^{-1}$\,K$^{-1}$ \\
Lower boundary pressure & 600 bar & Upper boundary pressure & $5\times10^{-6}$ bar \\
Horizontal resolution & C32 & Vertical resolution & 53 layers \\
Dynamical timestep & 25 s & Radiative timestep & 200 s \\
Drag timescale & $10^5$ s & & \\
\hline
\multicolumn{4}{l}{\textbf{2. 10$\times$ solar metallicity}} \\
\hline
Heat capacity & 12000 J\,kg$^{-1}$\,K$^{-1}$ & Specific gas constant & 3325.6 J\,kg$^{-1}$\,K$^{-1}$ \\
\hline
\multicolumn{4}{l}{\textbf{3. Cloud included}} \\
\hline
$q_{\rm deep}$ for MnS & $2.17\times10^{-5}$ kg\,kg$^{-1}$ & $q_{\rm deep}$ for Na$_2$S & $5.9\times10^{-5}$ kg\,kg$^{-1}$ \\
$\tau_c$ & 100 s & $\tau_{\rm deep}$ & 1000 s \\
\hline
\end{tabular}
\begin{tablenotes}
\footnotesize
\item Note: Three groups of simulations are conducted (control, 10$\times$ solar metallicity, and cloud), and each group includes both synchronous and nonsynchronous rotation cases. Default values follow the control group, and any modifications for each group are listed within the corresponding group.
\end{tablenotes}
\end{table*}

We prescribe a lognormal particle size distribution with a reference radius of 2 $\mu$m, and a nondimensional parameter that describes the width of the distribution $\sigma$ = 0.5. Then, based on the local mass mixing ratio of cloud tracers $q_c$ from Equation (\ref{eq6}), we can determine the particle size distribution, i.e., the number density as a function of particle radius. The wavelength-dependent extinction and scattering coefficients of clouds, which are spatially and time resolved, are then interpolated using pre-calculated Mie tables, summing over the cloud size distribution and were included in the radiative transfer calculation. A similar model and cloud setup have been applied to GCM modeling in \cite{bell2024} and \cite{biller2024}.

\subsection{Radiative transfer post-processing: PICASO}\label{sec:picaso}

PICASO is an open-source radiative transfer code for computing the reflected, thermal, and transmission spectra of planets and brown dwarfs \citep{batalha2019,mukherjee2023}. 
To simulate observable signals, we calculate phase-dependent thermal emission spectra using PICASO with wavelength bins of 196 from 0.26 to 227 $\mu$m (equivalent to a spectral resolution of $R\sim 100$ at the infrared) using the correlated-k technique, based on the 3D, time-dependent GCM outputs as in \cite{tan2024}. PICASO reads in the 3D thermal structure and calculates the intensity emerging from the top of the atmosphere at each longitude and latitude point. Then, it sums the 2D map intensity with appropriate weights to obtain a mean “observed” flux. This naturally captures the 3D information from the GCM. 
If clouds are included in the GCM, their time- and location-dependent opacity is included in the PICASO calculations.
We compute transmission spectra across 0.3–14 $\mu$m using the opacity-resampling scheme, which calculates the spectrum at a native spectral resolution of $R\sim10,000$ and then bin it down to $R\sim150$.
Since PICASO natively supports transmission spectra base on one-dimensional temperature-pressure profiles, we adapt the following procedure to take 3D GCM structures into account: 1) We define azimuth and zenith grids based on the star-planet-observer axis, following \cite{fortney2010} and \cite{macdonald2022}. The zenith angle is 0$^{\circ}$ at the substellar point, 90$^{\circ}$ at the terminator, and 180$^{\circ}$ at the antistellar point, while the azimuth angle is set to zero at the planetary north pole and increases toward the trailing hemisphere. 2) For simplicity, we average the 3D outputs (e.g., temperature, clouds) between 70$^{\circ}$--110$^{\circ}$ in zenith angle, since the transit signal primarily originates near the terminator. 3) We calculate the transit spectrum as a function of azimuth, and then average over azimuth to obtain the mean, wavelength-dependent transit depth. The corresponding mean P–T profile around the terminator is used at a particular azimuth angle.
PICASO and the ADAM share the same opacity database and equilibrium chemistry and this guarantees energetic consistency between the GCM and the radiative transfer post-processing.

\subsection{Experimental design and model paramters}\label{sec:experiments}

\begin{figure*}
    \centering
    \includegraphics[width=0.8\textwidth]{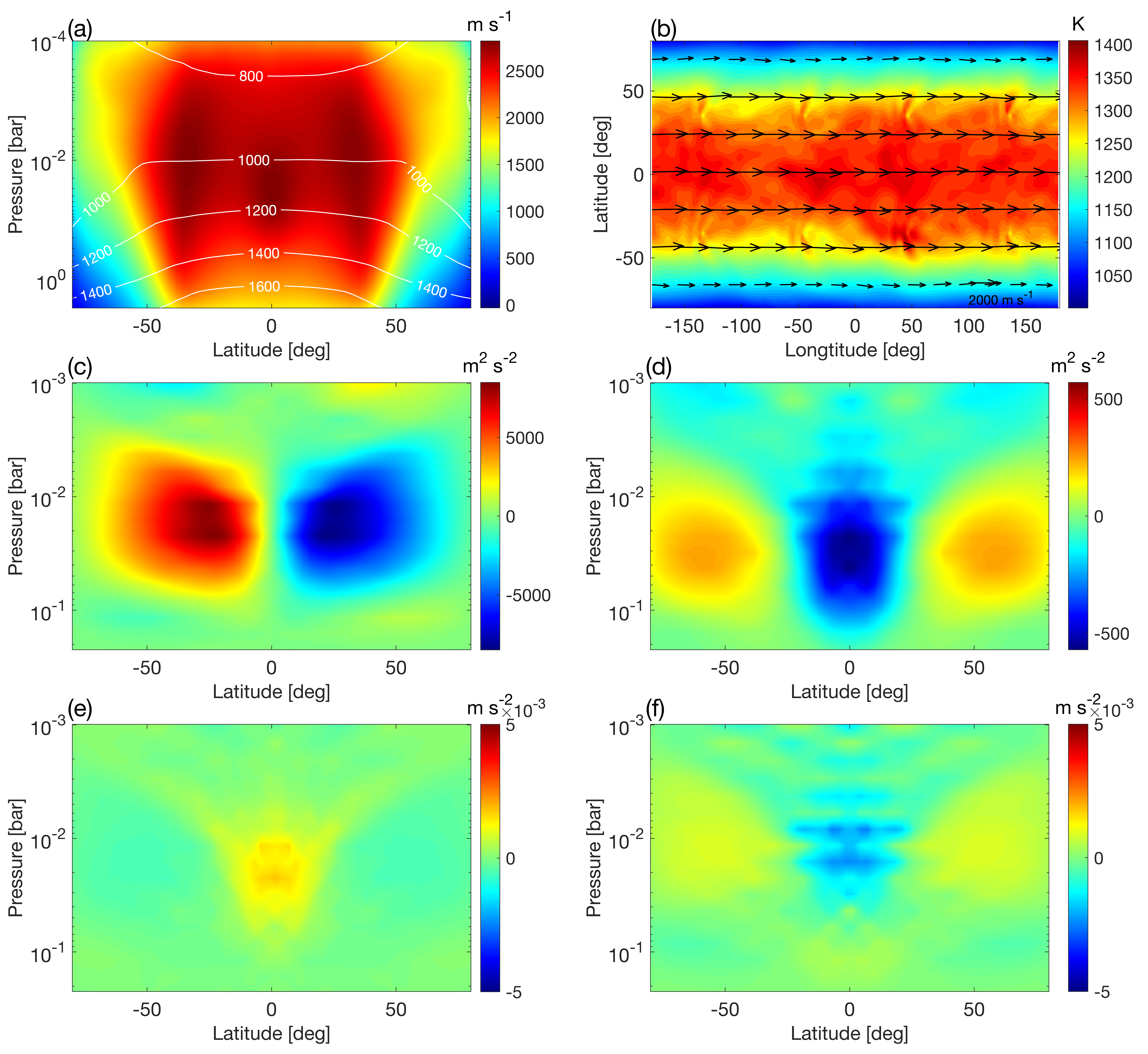}
    \caption{Atmospheric circulation pattern and mechanism of the formation of eastward jets in low- and mid-latitudes under synchronous rotation. (a) Zonal-mean zonal wind (shading, m\,s$^{-1}$) and temperature (contours, K) as functions of latitude and pressure. (b) Horizontal temperature map (K) near the photosphere ($p \sim 0.22$ bar), with horizontal winds indicated by arrows. 
    (c) Zonal-mean horizontal eddy momentum flux (m$^2$\,s$^{-2}$), with positive (negative) values indicating northward (southward) eddy momentum transport. (d) Zonal-mean vertical eddy momentum flux (m$^2$\,s$^{-2}$), with positive (negative) values implying upward (downward) eddy momentum transport; (e) Zonal-mean horizontal and (f) vertical eddy momentum convergence (m\,s$^{-2}$), with positive (negative) values implying eastward (westward) eddy acceleration. Diagnoses in panels (c-f) were calculated using GCM outputs during the spin-up phase, and  only values between $10^{-3}$ and $10^{-1}$ bar are shown, where the zonal jets are centered.}
    \label{fig:sync}
\end{figure*}

K2-290 b has the highest equilibrium temperature of 1230 K and is almost closest to Earth among the targets listed in Table \ref{tab:list}. Thus, we choose K2-290 b as a representative to investigate the circulation pattern and potential observational signals of high-obliquity mini-Neptunes. The planetary parameters of K2-290 b adopted in the simulations are listed in the bottom line of Table \ref{tab:list} \citep{hjorth2019}, including planet radius, gravity, semi-major axis, orbital period, and potential rotation period and planetary obliquity obtained from Equation~(\ref{eq:tce2}). Stellar parameters for GCM and post-processing are adopted from \cite{hjorth2019}. To explore the influence of high obliquity, we conduct two types of simulations: 1) the planet is synchronously rotating, with both rotation period and orbital period of 9.2 days, and with an obliquity of zero, 2) the planet is nonsynchronously rotating, with an obliquity of 67$^{\circ}$, a rotation period of 13.4 days, and an orbital period of 9.2 days.

No atmospheric measurement is available to constrain the atmospheric metallicity of K2-290 b, and we start out with default models with a solar metallicity and without clouds for circulation investigation following common GCM modeling strategies \citep{showman2020review}. Higher metallicity in the atmospheres of mini-Neptunes is certainly possible. Previous studies have shown that a higher metallicity could strengthen atmospheric winds by increasing gas opacity and then day-night contrast \citep{kataria2014,charnay2015,drummond2018}. 
The presence of clouds on extra-solar substellar atmospheres (e.g., hot-Jupiters, mini-Neptunes, brown dwarfs) has long been suggested by observations, which could strongly affect atmospheric opacity and the circulation pattern \citep[e.g.,][]{demory2013,kreidberg2014,helling2014,knutson2014,sing2016,benneke2019,feinstein2023}. Both atmospheric metallicity and clouds could potentially exert significant influences on circulation and observability of high-obliquity mini-Neptunes.
Two additional groups of simulations are conducted in this work: one with 10$\times$ solar metallicity without clouds and the other with MnS and Na$_2$S clouds at solar metallicity. Each group of simulations contains synchronously and nonsynchronously rotating cases. For cases with solar metallicity, the specific gas constant $R$ is 3714 J\,kg$^{-1}$\,K$^{-1}$ and heat capacity $c_p$ is 13000 J\,kg$^{-1}$\,K$^{-1}$. By comparison, for the cases with 10$\times$ solar metallicity, $R$ is 3325.6 J\,kg$^{-1}$\,K$^{-1}$ and $c_p$ is 12000 J\,kg$^{-1}$\,K$^{-1}$ \citep{charnay2015}.

\begin{figure*}
    \centering
    \includegraphics[width=0.8\textwidth]{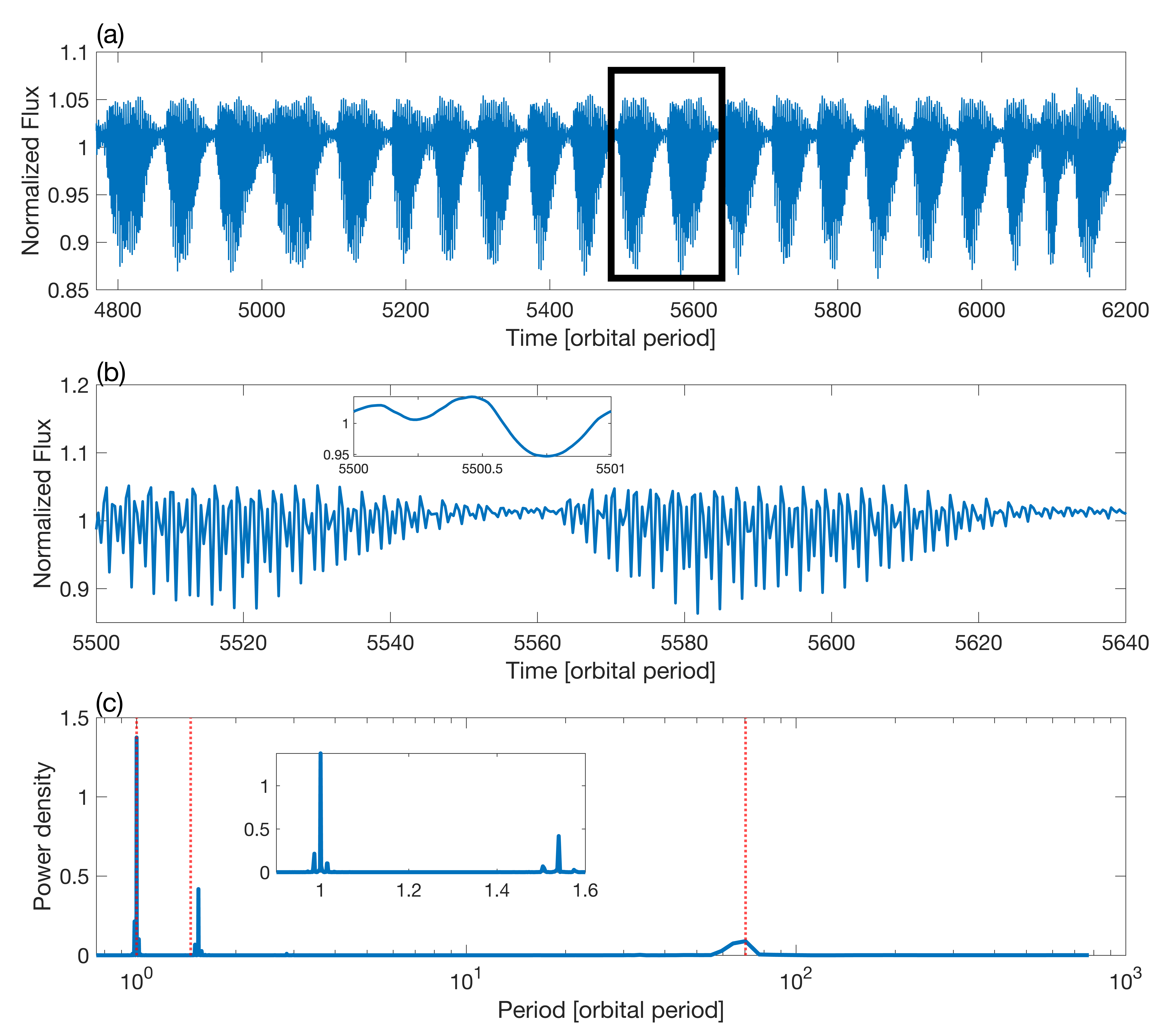}
    \caption{Temporal variability of normalized light curve from GCM simulations under nonsynchronous rotation. (a) Long-term normalized light curves viewed 30N-on as a function of simulation time over 4800$\sim$6200 orbits. (b) A zoom-in light curve between 5500 and 5640 orbital periods, and the temporal range is marked by the black rectangle in the top panel. The inset shows a zoom-in over a single orbit. (c) Power spectrum of periodogram for the normalized light curves. From left to right, the red dotted lines correspond to the orbital period, the rotation period, and the period of the long-term oscillations shown in the top panel ($\sim70$ orbital periods), respectively. The inset shows a magnified view near the orbital and rotation periods.}
    \label{fig:nonsync_normflux}
\end{figure*}

The horizontal resolution of the cubed sphere grid is C32 (equivalent to 128$\times$64 grid points in longitude and latitude), sufficient to resolve large-scale atmospheric structures given the slow rotation rate and small radius of high-obliquity mini-Neptunes in Table \ref{tab:list}. Vertically, the model has 53 layers evenly spaced in log pressure, which extends from 600 bar at the bottom and 5$\times$10$^{-6}$ bar at the top of the model. Temperatures at the bottom are relaxed towards a prescribed value of 2500 K, and this thermal bottom condition mimics the connection of the atmosphere to the interior of a given (assumed) entropy that does not evolve over short timescales \citep{komacek2022}.
The simulations are integrated from an initial rest state with a dynamical timestep of 25 s and a radiative timestep of 200 s. The experimental design and model parameters are summarized in Table \ref{tab:paras}.
For synchronous rotation cases, the results presented are time-averaged over hundreds of days once the models have reached statistical equilibrium.

\section{Results}\label{sec:result}

\subsection{Basic circulation pattern when synchronously rotating}\label{sec:sync}

When K2-290 b rotates synchronously, the circulation is dominated by globally eastward jets (Figure \ref{fig:sync}(a-b)), reaching peaking velocities of $\sim$2800 m\,s$^{-1}$ near the equator and $\pm30^{\circ}$ at $\sim$0.01 bar. These jets extend vertically from the top of the domain to $\sim$10 bar. The presence of strong winds in low- and mid-latitudes enables efficient longitudinal heat transport, resulting in a relatively weak horizontal temperature contrast---less than 100 K within a latitude range of $\sim\pm50^{\circ}$ near the photosphere (p$\sim$0.22 bar). 

Both the low- and mid-latitudes eastward jets exhibit superrotation, i.e., their angular momentum per mass exceeds the maximum angular momentum per mass of the underlying planet. The driving mechanism of equatorial superrotation on K2-290 b is analogous to that on typical warm/hot Jupiters and mini-Neptunes (e.g., \citealt{showman2011,showman2015,innes2022,landgren2023,liu2025}). Driven by day-night temperature contrast, standing planetary-scale waves transport eddy momentum from high-latitudes toward the equator, leading to eddy momentum convergence and the generation of equatorial superrotation (Figure \ref{fig:sync}(c \& e)). By comparison, the mid-latitude eastward jets are primarily accelerated by the vertical eddy momentum transport, which produces eddy momentum convergence in pressure layers above 10$^{-1}$ bar (Figure \ref{fig:sync}(d \& f)). The mean circulation also contributes to the generation of the mid-latitude jets (figure not shown).

The weak day-night temperature contrast shown in the model is attributed to the slow planetary rotation, relatively small planetary radius, and a moderate radiative timescale.
The effect of rotation on heat transport may be qualitatively assessed by the Weak-Temperature-Gradient (WTG) parameter $\wedge$ \citep{pierrehumbert2019}: 
\begin{equation}
    \wedge \equiv \frac{L_d}{a} = \frac{\sqrt{RT_{\rm eq}}}{\Omega a},
    \label{eq:WTGpara}
\end{equation}
where $L_d = \sqrt{RT_{\rm eq}}/\Omega$ is Rossby deformation radius, $R=R^{\star}/\mu$ is specific gas constant, $R^{\star}$ is the universal gas constant, $\mu = 2.2$ g\,mol$^{-1}$ is the mean molecular weight for a solar metallicity, $T_{\rm eq}$ is radiative equilibrium temperature, $a$ is planet radius, and $\Omega$ is planetary rotation rate. There is global WTG behaviour for $\wedge \gg 1$, significant horizontal temperature gradients for $\wedge \ll 1$, and WTG behaviour only at low latitudes for $\wedge \sim 1$ \citep{pierrehumbert2019}. For our simulated target K2-290 b, we obtain $\wedge \approx 14$, which lies in the WTG regime, indicating efficient horizontal heat transport and weak horizontal temperature gradients in the atmosphere, consistent with our GCM simulation. Note that the WTG argument applies to general irradiation conditions and is not limited to synchronous rotation.

In the synchronously rotating case, the characteristic day-night temperature contrast is additionally affected by the efficiency of radiative heating, more specifically, by the competition of the radiative timescale, wave propagation timescale, and planetary rotation timescale \citep{perez2013,komacek2016}. Applying the scaling theory of \cite{perez2013}, the characteristic day-night temperature difference relative to that at radiative equilibrium (denoted as a dimensionless variable $A$) can be estimated as, in a drag-free case: 
\begin{equation}
     A \sim (1+\frac{\tau_{\rm rad}}{\Omega\tau_{\rm wave}^2})^{-1},
     \label{eq:WTG}
\end{equation}
where $\tau_{\rm rad} = \frac{c_p p}{4g\sigma T_{eq}^3}$ is the radiative timescale, $p$ is the pressure near the photosphere, $\sigma$ is the Stephan-Boltzmann constant, $\tau_{\rm wave} = L/\sqrt{RT_{\rm eq}}$ is a gravity wave propagation timescale, and $L$ is a relevant length scale taken to be the planetary radius. $A\sim0$ indicates a nearly uniform atmospheric temperature due to efficient horizontal heat transport, whereas $A\sim1$ corresponds to a large day-night temperature contrast close to radiative equilibrium.
Given a moderate $T_{\rm eq}$ and radius of K2-290 b (Table \ref{tab:list}), and a photospheric pressure of $\sim$0.22 bar, we obtain $\tau_{\rm rad}\sim 6\times10^4$ s, $\tau_{\rm wave}\sim 10^4$ s, and $A \approx 0.02$, also consistent with the weak day-to-night temperature gradient shown in our GCM results.

The slow rotation rate of K2-290 b is also responsible for the meridionally broad equatorial jet shown in Figure \ref{fig:sync}(a-b). The characteristic width of the equatorial jet driven by the day-night forcing is proportional to the equatorial Rossby deformation radius, $L_{\rm eq} = (\sqrt{RT}a/2\Omega)^{1/2}$, in an equatorial $\beta$-plane approximation \citep[e.g.,][]{showman2011,tan2020}. Here, the $\beta$-plane assumes the Coriolis parameter varies linearly with latitude, i.e., $f=\beta y$, where $\beta=2\Omega/a$ is the meridional derivative of the Coriolis parameter at the equator and $y$ is the northward distance \citep[e.g.,][]{vallis2017atmospheric}. Since the equatorial Rossby deformation radius is inversely proportional to rotation rate, the equatorial jet becomes wider as rotation rate decreases. For K2-290 b, $L_{\rm eq}\sim 5\times10^7$ m, larger than the planetary radius, indicating that large-scale equatorial wave dynamics dominate the global circulation while the Coriolis force plays a relatively minor role \citep{pierrehumbert2019,showman2020review}. Although the above argument in the $\beta$-plane does not formally hold when $L_{\rm eq}> a$, simulations of tidally locked GCMs with slow rotation rates showed that as the rotation rate decreases, the equatorial superrotating jet widens and eventually extends to a global scale \citep[e.g.,][]{beltz2020,showman2015,haqq2018demarcating,lewis2021}.  

\begin{figure*}
    \centering
    \includegraphics[width=0.8\textwidth]{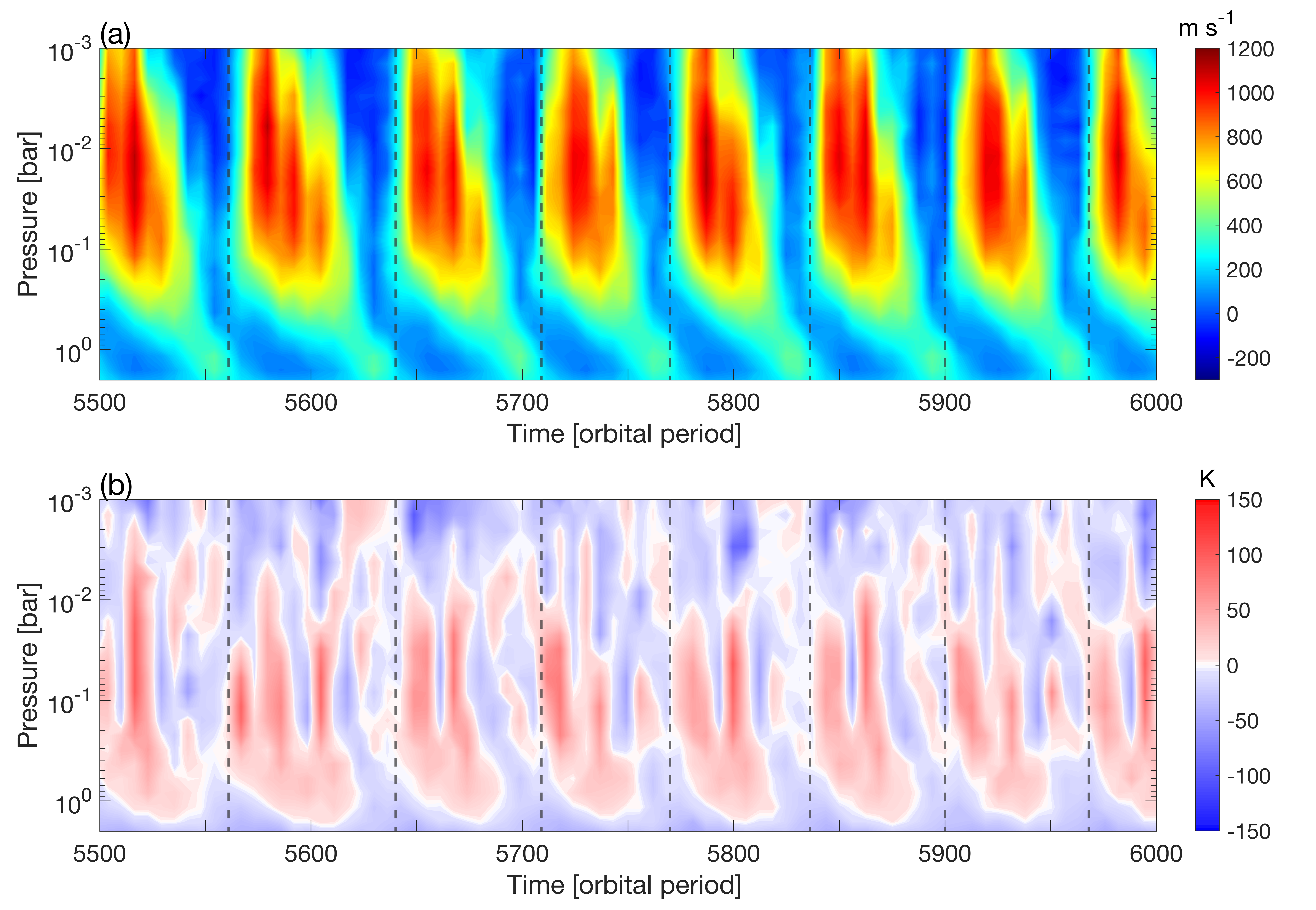}
    \caption{QBO-like oscillations under nonsynchronous rotation. (a) Timeseries of zonal-mean zonal wind at 1.41$^{\circ}$N as functions of pressure. (b) Timeseries of zonal-mean equator-to-pole temperature contrast. Positive/negative values of the equator-to-pole temperature contrast indicate higher/lower equatorial than polar temperatures. The onset time of each QBO-like oscillation is marked with dashed lines in both panels. }
    \label{fig:nonsync_utime}
\end{figure*}

\subsection{Basic circulation pattern when nonsynchronously rotating}\label{sec:nonsync}

\subsubsection{Seasonal and long-term variations}\label{sec:nonsync_season}

When K2-290 b rotates nonsynchronously (with an obliquity of 67$^{\circ}$, a rotation period of 13.4 days, and an orbital period of 9.2 days), the high obliquity causes the substellar latitude to shift between 67$^{\circ}$N (summer solstice) to 67$^{\circ}$S (winter solstice) over the orbital timescale, resulting in significant variations in latitudinal insolation distribution. Specifically, the polar regions receive more insolation than the equator during solstices, while the equator receives more insolation than the poles during equinoxes. These variations induce seasonal variations in circulation patterns. Crucially, when the planetary obliquity exceeds 54$^{\circ}$, the annual-mean insolation is greater at the poles than at the equator \citep{ward1974,ferreira2014}, potentially reversing the meridional temperature gradient and associated wind patterns compared to low-obliquity planets.

\begin{figure*}
    \centering
    \includegraphics[width=0.9\textwidth]{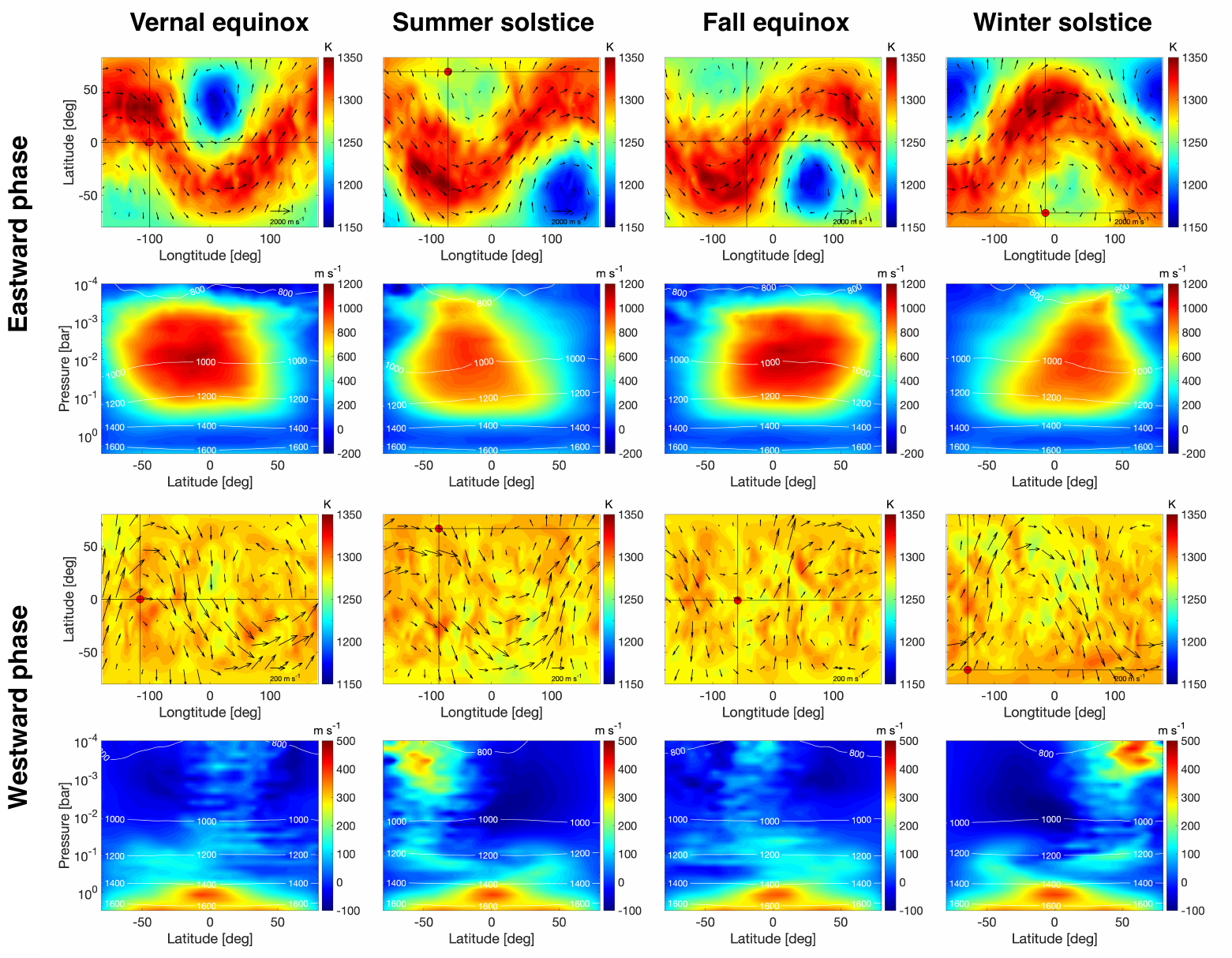}
    \caption{Seasonal variations of circulation patterns during the eastward phase (the top two rows) and the westward phase (the bottom two rows) under nonsynchronous rotation. Each row, from left to right, corresponds to the snapshot at vernal equinox, summer solstice, fall equinox, and winter solstice. For each phase: the top panels show horizontal temperature maps (K) near the photosphere ($\sim$0.2 bar) with horizontal winds as arrows as functions of longitude and latitude. The intersection of the horizontal and vertical black lines (marked by red filled circles) denotes the substellar point. The bottom panels display zonal-mean zonal wind (colors, m\,s$^{-1}$) and temperature (contours, K) as functions of latitude and pressure. }
    \label{fig:nonsync_4seasons}
\end{figure*}

The atmospheric circulation naturally exhibits a seasonal cycle driven by the non-zero planetary obliquity. In addition, the nonsynchronous simulation shows a long-term variability over a characteristic period of $\sim$70 orbital periods, likely a result of internal dynamical interactions. Before discussing the detailed circulation patterns, we summarize the seasonal and long-term atmospheric variations by showing a rotational light curve from the GCM in Figure \ref{fig:nonsync_normflux}.
Figure \ref{fig:nonsync_normflux}(a) shows the simulated light curves viewed 30N-on\footnote{Viewed by a hypothetical observer, the angle between the line-of-sight vector and the planetary spin vector can lie between $90^{\circ}-\phi$ and $90^{\circ}+\phi$ with a given planetary obliquity $\phi$ and assuming that the precession of the spin axis is negligible over a timescale of our interest. In Figure \ref{fig:nonsync_normflux}, we choose a viewing angle of 30N-on for demonstration, corresponding to an angle of 60$^\circ$ between the planetary spin axis and the line of sight. Light curves viewed equator-on (90$^{\circ}$) maximize the {\it rotational} variations and those viewed more pole-on maximize {\it seasonal} variations.} , which is obtained from integrating the top-of-atmosphere bolometric thermal flux of the GCM as the planet rotates. Figure \ref{fig:nonsync_normflux}(b) is a zoom-in in time for a closer inspection. The quasi-periodic long-term evolution with a period of about 70-80 orbital periods is the dominant feature in Figure \ref{fig:nonsync_normflux}(a). Inside each light-curve ``envelope" are the high-frequency variations corresponding to orbital (seasonal) and rotational variations  in the middle panel. The overall light-curve amplitude is only about 10$\%$, a small value in the context of close-in gas giants \citep{parmentier2018}. This small amplitude is because of the weak horizontal temperature variations as a result of the WTG regime for this planet, as discussed in section \ref{sec:sync}.

To quantify the periodicities, we perform a periodogram analysis of the light curve covering over 1500 orbital periods and obtain the power spectrum as a function of period (Figure \ref{fig:nonsync_normflux}(c)). Three major peaks appear at the orbital period, rotation period, and about 70 orbital periods, as expected.

The long-term variability is induced by the quasi-periodic oscillation in the wind and temperature patterns. 
Figure \ref{fig:nonsync_utime} shows the timeseries of zonal-mean zonal wind near the equator and zonal-mean equator-to-pole temperature difference under nonsynchronous rotation. The results exhibit alternating eastward and westward jets with an oscillation period of 70-80 orbital periods ($\sim$2 Earth years), accompanied by positive and negative equator-to-pole temperature contrasts in pressure layers between $10^{-2}$ and $10^{0}$ bar, respectively. For clarity, we hereafter refer to the state with eastward jets and higher equatorial temperatures as the eastward phase, and the state with westward jets and lower equatorial temperatures is the westward phase.
During the eastward phase, the zonal jets peak at $\sim$1200 m\,s$^{-1}$ near 0.01 bar, accompanied by a maximum equator-to-pole temperature contrast of $\sim$100 K near 0.1 bar. Notably, the eastward jets exhibit a downward migration in pressure levels between 0.1 and 1 bar. In the westward phase, the zonal jets reach about $-200$ m\,s$^{-1}$ at the same level, with a minimum temperature contrast of $-50$ K near 0.1 bar.

\begin{figure*}
    \centering
    \includegraphics[width=0.9\textwidth]{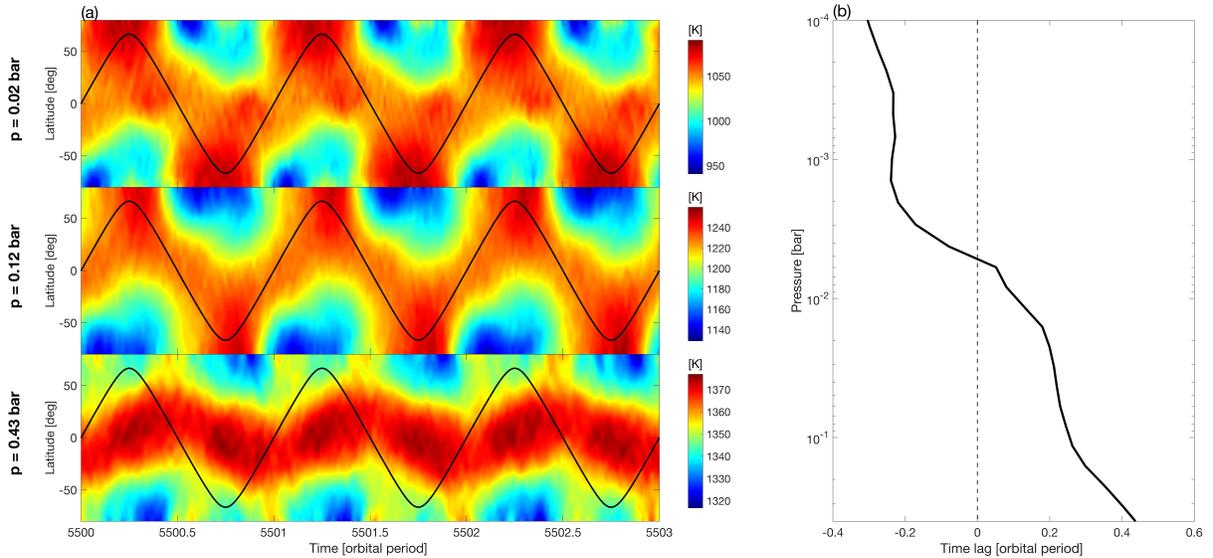}
    \caption{(a) Timeseries of zonal-mean temperature and substellar latitude versus latitude within three orbital periods at pressures of 0.02 bar (top), 0.12 bar (middle), and 0.43 bar (bottom) during the eastward phase, with the timeseries of substellar latitude indicated by black solid lines. (b) Time lag between the latitude of maximum zonal-mean temperature and the substellar latitude as a function of pressure, with zero time lag indicated by a dashed line. Positive values indicate that the temperature lags behind the insolation, while negative values indicate that it leads.}
    \label{fig:nonsync_ttime}
\end{figure*}

These oscillations resemble the quasi-biennial oscillation (QBO) in Earth's stratosphere, which features alternating equatorial zonal jets and temperature variations that migrate from low to high pressure with a period of $\sim28$ months \citep{baldwin2001}. 
The physical mechanism driving the QBO-type oscillations will be analyzed in Section \ref{sec:nonsync_qbo_mecha}. The comparison between Figure \ref{fig:nonsync_normflux} and Figure \ref{fig:nonsync_utime} shows that the temporal evolutions of the circulation and temperature patterns are highly correlated with the light curve. In our QBO-like oscillations, the light curve amplitude is high during the eastward phase, due to the relatively higher equator-to-pole temperature variations, whereas it is low during the westward phase, consistent with the relatively lower equator-to-pole temperature variations. 
The coupling between zonal wind and temperature patterns may be understood by the thermal wind balance, which suggests that a warmer (cooler) equator relative to the poles corresponds to an eastward (westward) jet near the photosphere  \citep{andrews1987,vallis2017atmospheric}. 

The atmospheric circulation exhibits distinct seasonal variations between the eastward and westward phases. Figure \ref{fig:nonsync_4seasons} shows the circulation and temperature patterns from vernal equinox to winter solstice for both phases. 
During the eastward phase, the photospheric temperature pattern is dominated by a strong hemispheric-scale cold vortex, with a minimum temperature of $\sim$1150 K in the winter hemisphere, and a weaker cold center in the summer hemisphere (the topmost row of Figure \ref{fig:nonsync_4seasons}). 
From summer solstice to winter solstice, the stronger cold vortex migrates from $\sim$50$^{\circ}$S to $\sim$50$^{\circ}$N over half an orbital period, while temperatures outside these two cold vortexes remain relatively high and uniform. Consequently, the horizontal pressure gradients induced by the cold vortices drive meandering eastward jets that extend from the equator to high latitudes throughout the year (the top two rows of Figure \ref{fig:nonsync_4seasons}). 
In turn, these strong zonal jets help sustain the cold vortices. On slowly rotating planets such as K2-290 b and Venus, the strong jets can establish cyclostrophic balance, i.e., a balance between horizontal centrifugal and latitudinal pressure gradient force, which maintains evident low-temperature regions \citep{showman2010,Bengtsson2012Venus}.

During the westward phase, the distinct cold vortex and continuous hot region near the photosphere disappear, possibly due to the absence of strong zonal jets (the bottom two rows of Figure \ref{fig:nonsync_4seasons}). 
Instead, the temperature field becomes more homogeneous over the global scale but turbulent in smaller scales, with horizontal contrast less than 60 K near the photosphere. Roughly speaking, colder temperatures remain concentrated in the winter hemisphere during solstices. Zonal-mean zonal wind above 10$^{-1}$ bar shows a more even distribution between eastward and westward components, with equatorial winds oscillating seasonally in response to the meridional distribution of insolation (the bottommost rows of Figure \ref{fig:nonsync_4seasons}). Below 10$^{-1}$ bar, zonal-mean zonal winds remain predominantly eastward.

\begin{figure*}
    \centering
    \includegraphics[width=0.8\textwidth]{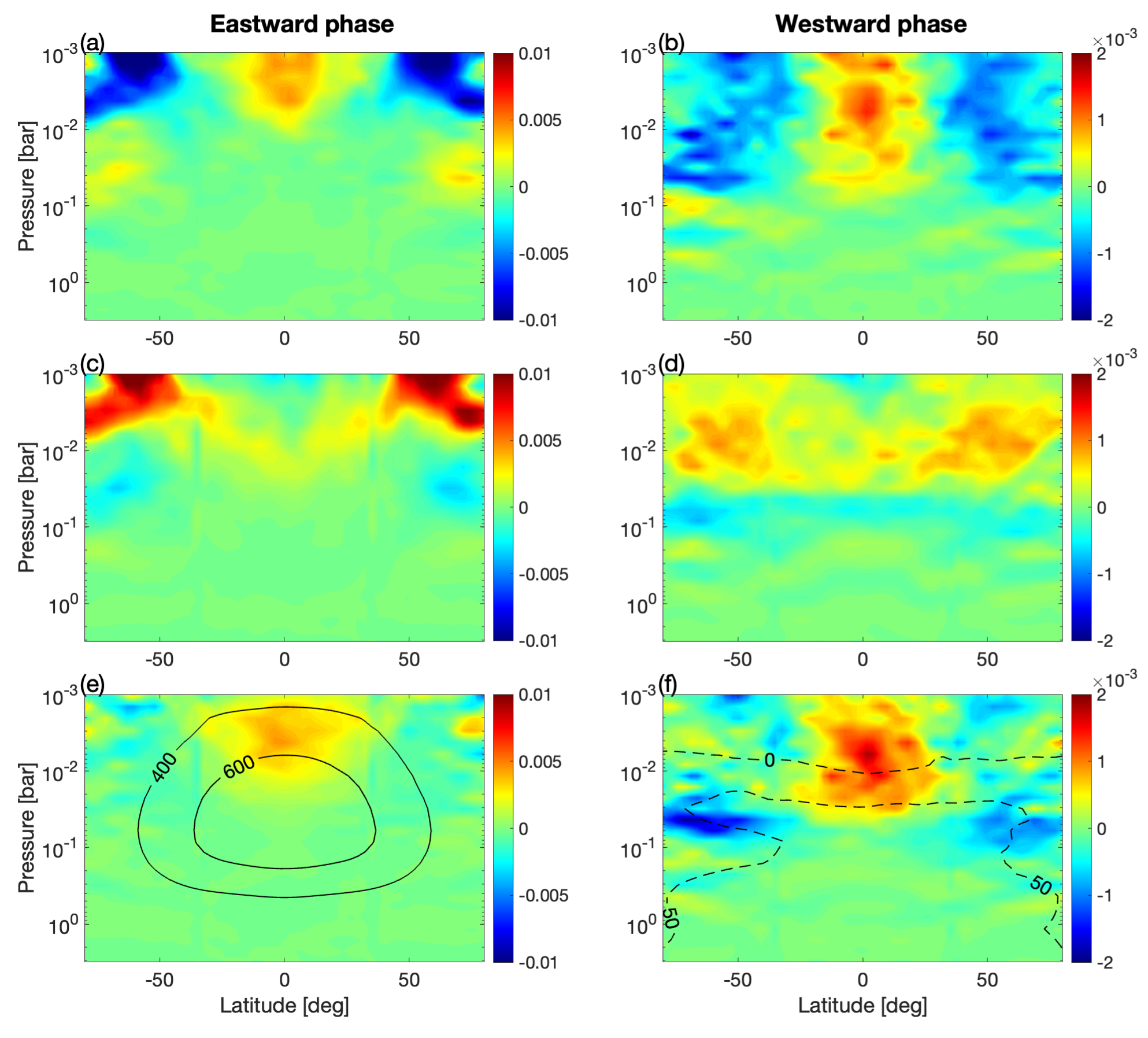}
    \caption{Zonal-mean Eliassen–Palm (EP) flux divergence (colors, m\,s$^{-2}$) during the eastward (left panels) and westward phases (right panels) under nonsynchronous rotation. From top to bottom panels: horizontal eddy momentum divergence (a \& b), vertical eddy momentum divergence (c \& d), and total eddy momentum divergence (e \& f). Positive (negative) values correspond to eastward (westward) accelerations to the zonal-mean zonal wind. In the bottom panels, zonal-mean zonal wind is overlaid(contours, m\,s$^{-1}$), with solid lines denoting positive values and dashed lines denoting negative values. The results are averaged over one orbital period, such that seasonal variations are filtered out.}
    \label{fig:nonsync_EP}
\end{figure*}

There is a clear phase lag in the zonal-mean temperature relative to the stellar irradiation, particularly at high pressures. As shown in Figure \ref{fig:nonsync_ttime}, the time lag between the latitude of maximum zonal-mean temperature and the substellar latitude increases with pressure, reaching approximately 0.4 orbital periods (corresponding to about $70^{\circ}$) at 0.4 bar.
The amplitude and phase lag of the zonal-mean temperature near the photospheric region can be understood in terms of the ratio between the orbital timescale ($P_{\rm orb}$ = 9.2 days) and the radiative timescale ($\tau_{\rm rad}$). A small $P_{\rm orb}/\tau_{\rm rad}$ results in small amplitudes and a large phase lag of temperature relative to irradiation, and vice versa \citep[e.g.,][]{tan2022}. The radiative timescale generally increases with pressure, leading to weak seasonal variations and an evident phase lag in deep layers. 
Notably, at pressure smaller than $\sim$0.005 bar, the time lag becomes negative, indicating that the temperature maximum leads the stellar irradiation (Figure \ref{fig:nonsync_ttime}(b)). In this low-pressure region, the shortwave radiative flux is limited (see Section \ref{sec:10solar_cloud} and Figure \ref{fig:flux_compare}), such that the temperature evolution is controlled primarily by dynamical processes rather than direct radiative forcing.

\subsubsection{Mechanism of the QBO-like oscillations}\label{sec:nonsync_qbo_mecha}

Classical theories of QBO-like oscillations suggest that the alternating eastward and westward jets result from the interaction of the upward propagating waves and the background mean flow \citep{lindzen1968,holton1972,baldwin2001}. Here we apply diagnoses similar to \cite{showman2019}, showing that the QBO-type oscillation is driven by similar mechanisms. To determine the eddy momentum fluxes in the wave-mean-flow interactions, we refer to the Eliassen–Palm (EP) flux equation in pressure coordinates \citep{andrews1983}:
\begin{equation}
\begin{aligned}
    F &= (F_{\phi}, F_p) \\ 
      &= acos\phi\lbrace{-\overline{u'v'}+\frac{\overline{v'\theta'}}{\partial \overline{\theta}/\partial p}}, \overline{u'\omega'} - \frac{\overline{v'\theta'}}{\partial \overline{\theta}/\partial p}\lbrack{\frac{1}{acos\phi}}\frac{\partial(\overline{u}cos\phi)}{\partial \phi}-f \rbrack \rbrace.
    \label{eq:epflux}
\end{aligned}
\end{equation}
Here $a$ is planet radius, $\phi$ is latitude, $f=2\Omega sin\phi$ is the Coriolis parameter, and $\Omega$ is the planetary rotation rate. 
Overbars represent zonal means and primes represent deviations from zonal means. 
The divergence of the EP flux is:
\begin{equation}\label{eq:epdiv}
    \nabla \cdot F = \frac{1}{acos\phi}\frac{\partial (F_{\phi}cos\phi)}{\partial \phi} + \frac{\partial F_p}{\partial p},
\end{equation}
where positive divergence ($\nabla \cdot F > 0$) corresponds to eastward (positive) eddy-induced acceleration to the zonal-mean zonal wind, and negative divergence corresponds to westward acceleration.

\begin{figure}
    \centering
    \includegraphics[width=0.9\linewidth]{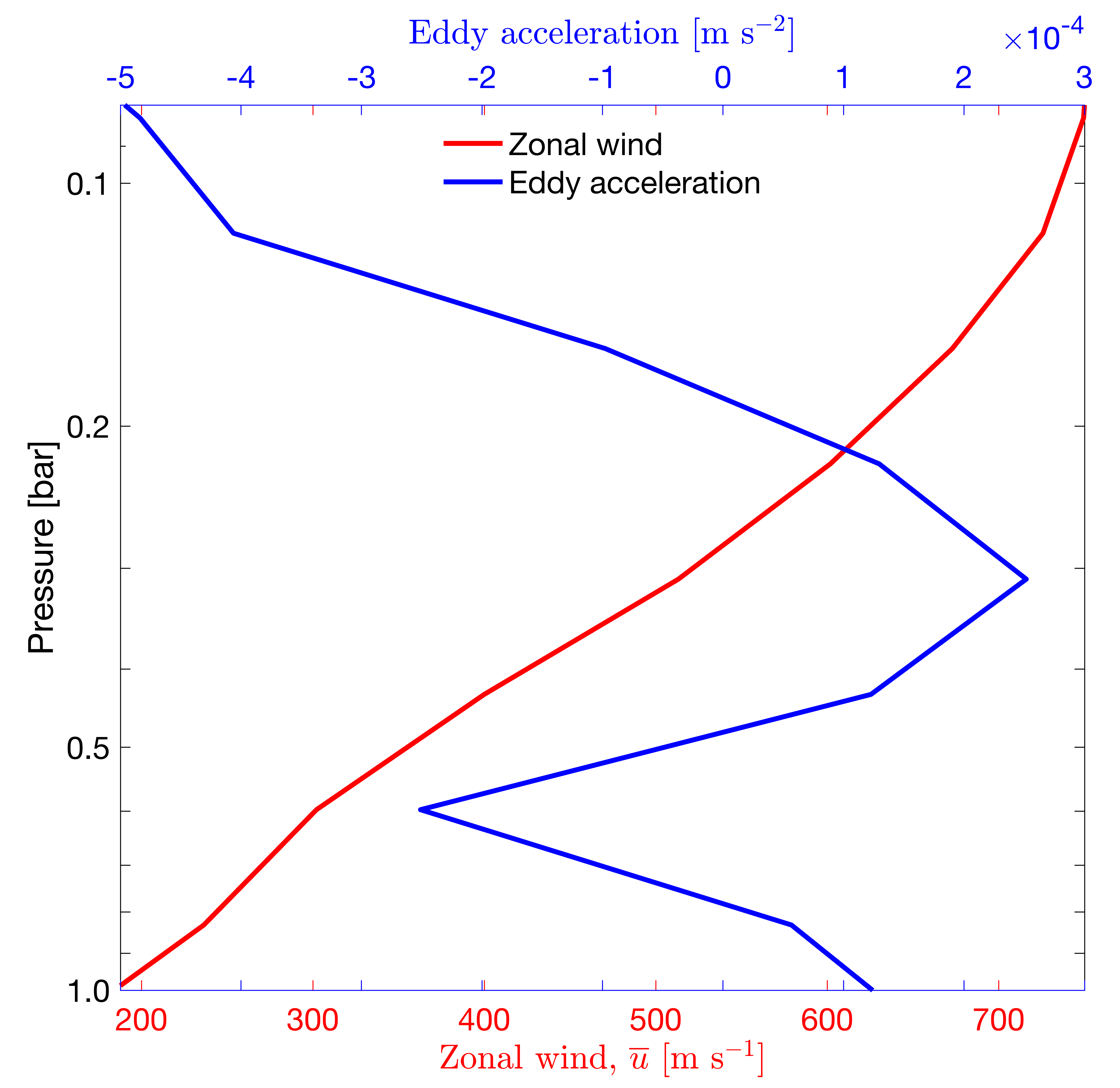}
    \caption{Zonal- and equatorial-mean (within $\pm10^{\circ}$ latitude) eddy acceleration from Equation (\ref{eq:epdiv}) (blue, m\,s$^{-2}$) and zonal wind (red, m\,s$^{-1}$) as functions of pressure during the eastward phase of the QBO-like oscillations under nonsynchronous rotation. Only values between 0.1 and 1 bar are shown, where the downward migration of eastward jets is most evident. The results are averaged over one orbital period to filter out seasonal variations.}
    \label{fig:nonsync_EP2}
\end{figure}

The vertical transport of eddy momentum plays a key role in the formation of QBO-like oscillations. The zonal-mean flow at pressures lower than $\sim0.1$ bar exhibits little downward migration but significant periodic oscillations, likely caused by the change of eddy momentum fluxes.  Figure \ref{fig:nonsync_EP} shows the horizontal divergence, vertical divergence, and total divergence of the EP flux during the eastward and westward phases. In the eastward phase (Figure \ref{fig:nonsync_EP}(a,c,e)), both horizontal and vertical eddy momentum convergence contribute to eastward acceleration within $\pm$30$^{\circ}$ latitude, supporting the persistence of eastward jets. In the westward phase (Figure \ref{fig:nonsync_EP}(b,d,f)), the horizontal eddy momentum still favors eastward acceleration, but with a magnitude roughly one order smaller than that in the eastward phase. 
However, the vertical component exhibits a vertically stacked pattern of eastward and westward accelerations, leading to vertically stacked zonal jets (the bottommost panels of Figure \ref{fig:nonsync_4seasons}).

At pressures between about 0.1 and 1 bar, the zonal-mean flow shows clear downward migration patterns during the eastward phase of the periodic oscillations, resembling a more classical QBO-like behavior (Figure \ref{fig:nonsync_utime}(a)). No such propagation is observed during the westward phase, in contrast to QBO-like oscillations in previous studies \citep{baldwin2001,showman2019,lian2023}. Figure \ref{fig:nonsync_EP2} shows the vertical profiles of equatorial-mean eddy-induced acceleration and zonal wind between 0.1 and 1 bar during the eastward phase. The maximum eastward (positive) acceleration occurs on the lower flank of the eastward jet. Near the equator where the balancing Coriolis force associated with the mean meridional circulation to the eddy forcing is weak, the eddy forcing causes the downward migration of the jets. 
Upward-propagating waves generated in the lower atmosphere have both eastward and westward phase speeds. In the presence of a background zonal flow, they are absorbed at critical layers where the zonal jet speed equals the wave phase speed, inducing eddy accelerations \citep{lindzen1968,holton1972}. Consequently, eastward-propagating waves are absorbed near the base of the eastward jet, inducing eastward acceleration. This wave-mean flow interaction results in the downward propagation of eastward jets.

\begin{figure*}
    \centering
    \includegraphics[width=0.85\textwidth]{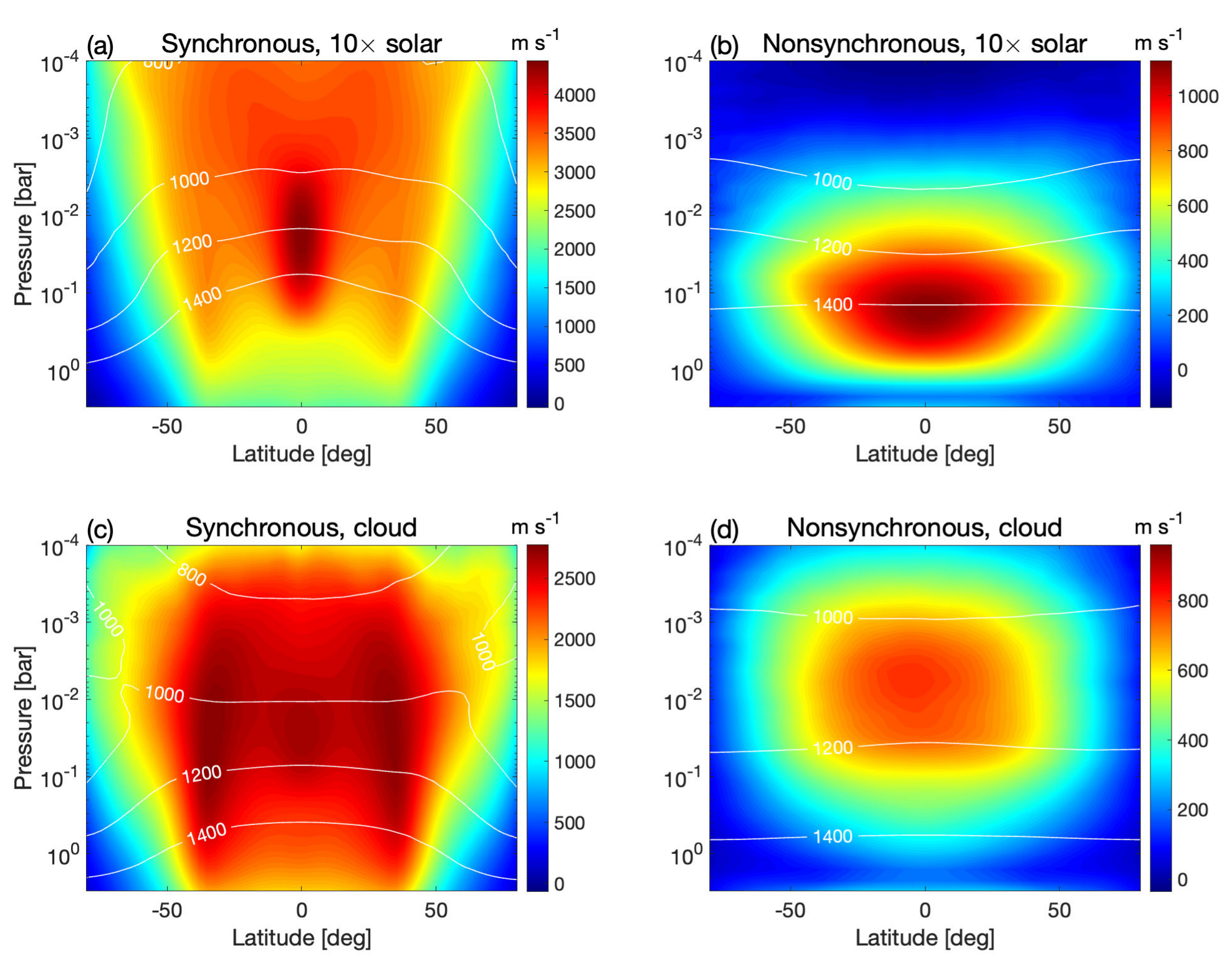}
    \caption{Influence of higher metallicity and clouds on atmospheric circulation and temperature patterns under synchronous (left) and nonsynchronous (right) rotation. Top panels: Zonal-mean zonal jets (colors, m\,s$^{-1}$) and temperatures (contours, K) with 10$\times$ solar metallicity. Bottom panels: corresponding results with clouds included. For the nonsynchronous cases, one-orbit-mean patterns during the eastward phase are shown. }
    \label{fig:influence_10solar_cloud}
\end{figure*}

It is worth noting that these QBO-type oscillations occur from low- to mid-latitudes because of the relatively slow rotation rate of K2-290 b. By comparison, QBO-type oscillations on relatively fast-rotating planets, including Earth, Jupiter, and brown dwarfs, are generally confined closer to the equatorial region, as eddy-induced accelerations are largely canceled by Coriolis forces in the extratropics \citep{baldwin2001,showman2019,tan2022bd,lian2023}. 
It should be noted that neither high obliquity nor asynchronous rotation alone can trigger QBO-type oscillations (Appendix \ref{ap:2cases}). Instead, such oscillations arise only within specific parameter spaces where upward-propagating waves are generated and interact with the background mean flow to drive QBO-like behaviors.

\begin{figure*}
    \centering
    \includegraphics[width=0.8\textwidth]{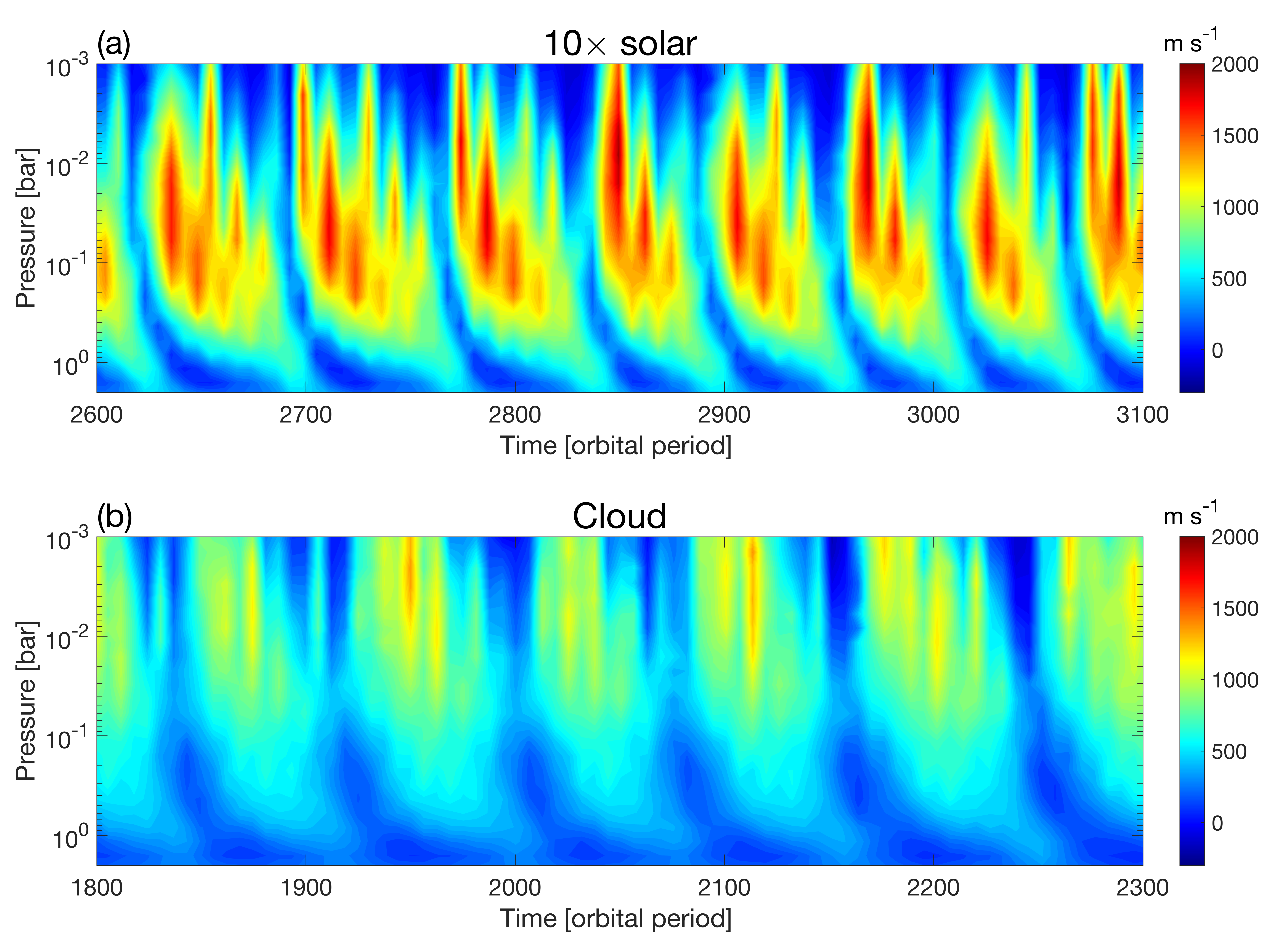}
    \caption{Timeseries of zonal-mean zonal wind at 1.41$^{\circ}$N as functions of pressure for the nonsynchronous cases with cloud-free and 10$\times$ solar metallicity (a) and with clouds included and 1$\times$ solar metallicity (b). QBO-like oscillations occur in both cases under nonsycnhronous rotation. }
    \label{fig:10solar_cloud_qbo}
\end{figure*}

\subsection{Higher Metallicity and Including Clouds}\label{sec:10solar_cloud}

\subsubsection{Influence of higher metallicity}\label{sec:nonsync_metal}

\begin{figure}
    \centering
    \includegraphics[width=0.9\linewidth]{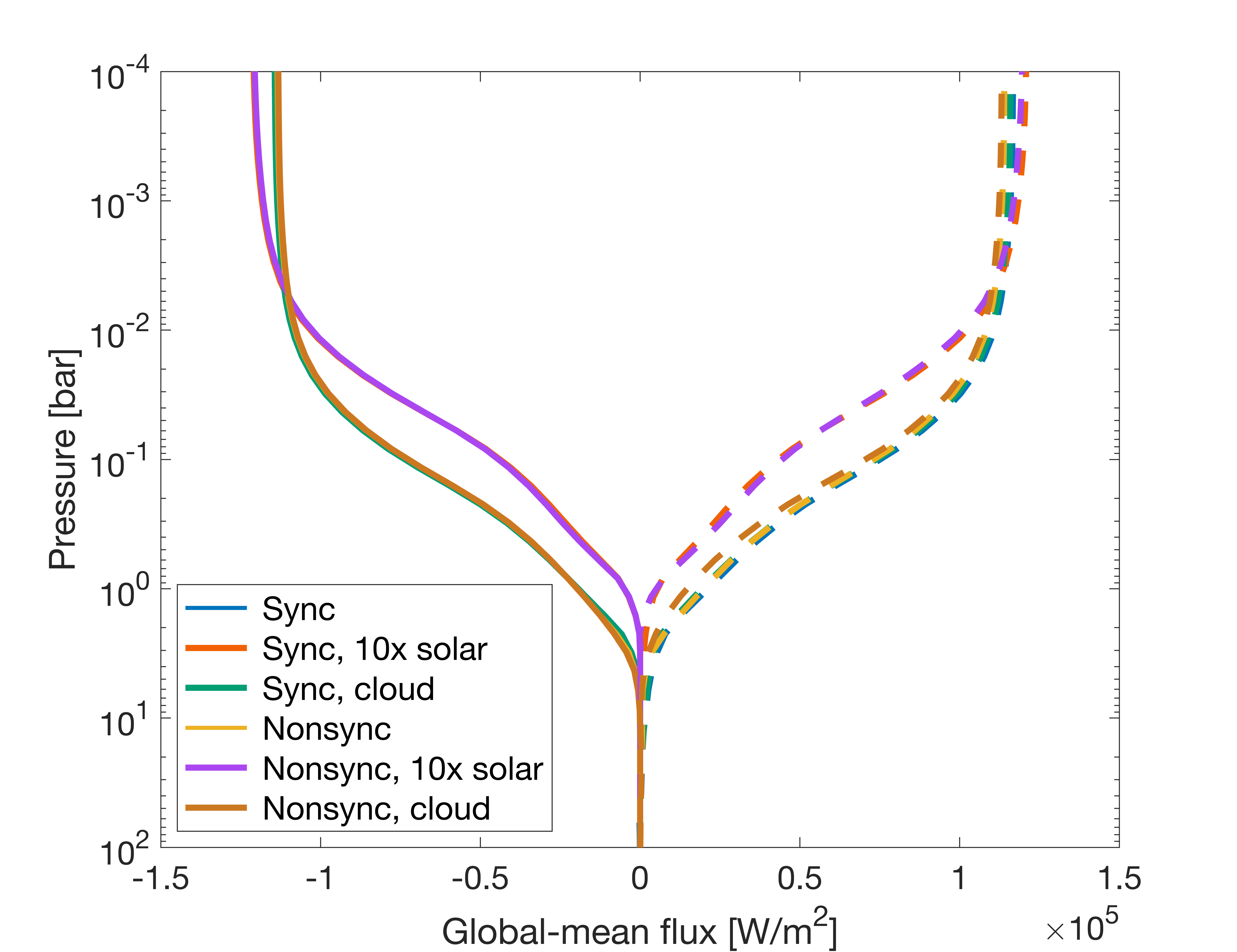}
    \caption{Comparison of global-mean shortwave (solid) and longwave radiative fluxes (dashed) between different cases: (i) synchronous rotation (blue), (ii) synchronous rotation with 10$\times$ solar metallicity (orange), (iii) synchronous rotation with clouds (green), (iv) nonsynchronous rotation (yellow), (v) nonsynchronous rotation with 10$\times$ solar metallicity (purple), and (vi) nonsynchronous rotation with clouds (brown). Positive values correspond to upward longwave flux, while negative values correspond to downward shortwave flux. The orange and purple lines are nearly identical, as are the other colored lines.  This indicates that shortwave absorption occurs primarily between 0.01 and 3 bar under solar metallicity and shifts slightly upward to lower pressures at 10$\times$ solar metallicity. For the nonsynchronous cases, one-orbit-mean values during the eastward phase is shown. }
    \label{fig:flux_compare}
\end{figure}

Previous studies suggest that higher atmospheric metallicity may significantly influence the atmospheric circulation \citep{kataria2014,charnay2015,zhang2017,drummond2018}, which may in turn affect the quantitative temperature variation and observational signatures of high-obliquity mini-Neptunes. 
Here, we investigate the influence of higher metallicity by performing two simulations with 10$\times$ solar metallicity under synchronous and nonsynchronous rotation conditions while holding all other parameters unchanged, except the specific gas constant and heat capacity (Table~\ref{tab:paras}). As shown in Figure \ref{fig:influence_10solar_cloud}(a-b), the zonal jets strengthen with higher metallicity. Under synchronous rotation, the peak equatorial jet speed reaches 4300 m\,s$^{-1}$. In the nonsynchronous case, the QBO-type oscillations still occur, while the peak eastward jet speed reaches 2000 m\,s$^{-1}$ (Figure \ref{fig:10solar_cloud_qbo}(a)). In addition, the jet center shift deeper in pressure under higher metallicity.

\begin{figure}
    \centering
    \includegraphics[width=0.9\linewidth]{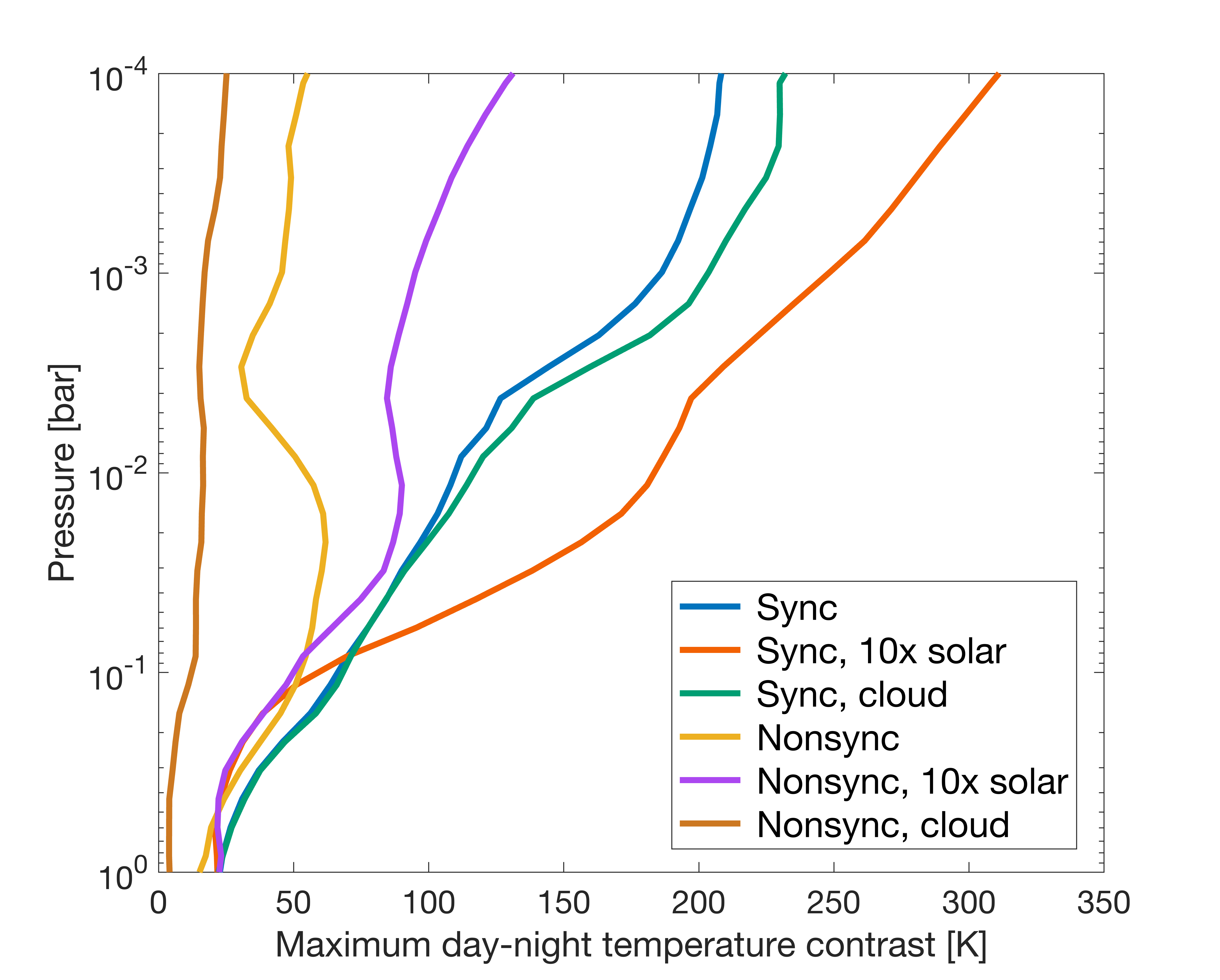}
    \caption{One-orbit-mean and latitudinal weighted maximum zonal temperature contrast as a function of pressure for different cases.}
    \label{fig:nonsync_tempdiff}
\end{figure}

Generally, higher metallicity enhances the abundance of strongly absorbing species, resulting in larger gas opacity. As a consequence, the incident stellar insolation is absorbed at lower pressures, leading to shallower heating and a lower-pressure photosphere (Figure \ref{fig:flux_compare}). In our case, the pressure level corresponding to the radiative equilibrium temperature (1230 K) shifts from $\sim$0.2 bar to $\sim$0.01 bar (Figure \ref{fig:influence_10solar_cloud}(a-b)). This transition shortens the radiative timescale, thereby strengthening the radiative forcing. 
Consequently, horizontal temperature contrasts increase (Figure \ref{fig:nonsync_tempdiff}), which then induces stronger eddy momentum convergence and more intense zonal jets. 
Under nonsynchronous rotation, similarly, zonal winds are also overall stronger than those in the 1$\times$ solar metallicity (Figure \ref{fig:10solar_cloud_qbo}(a)). 

\begin{figure*}
    \centering
    \includegraphics[width=0.7\textwidth]{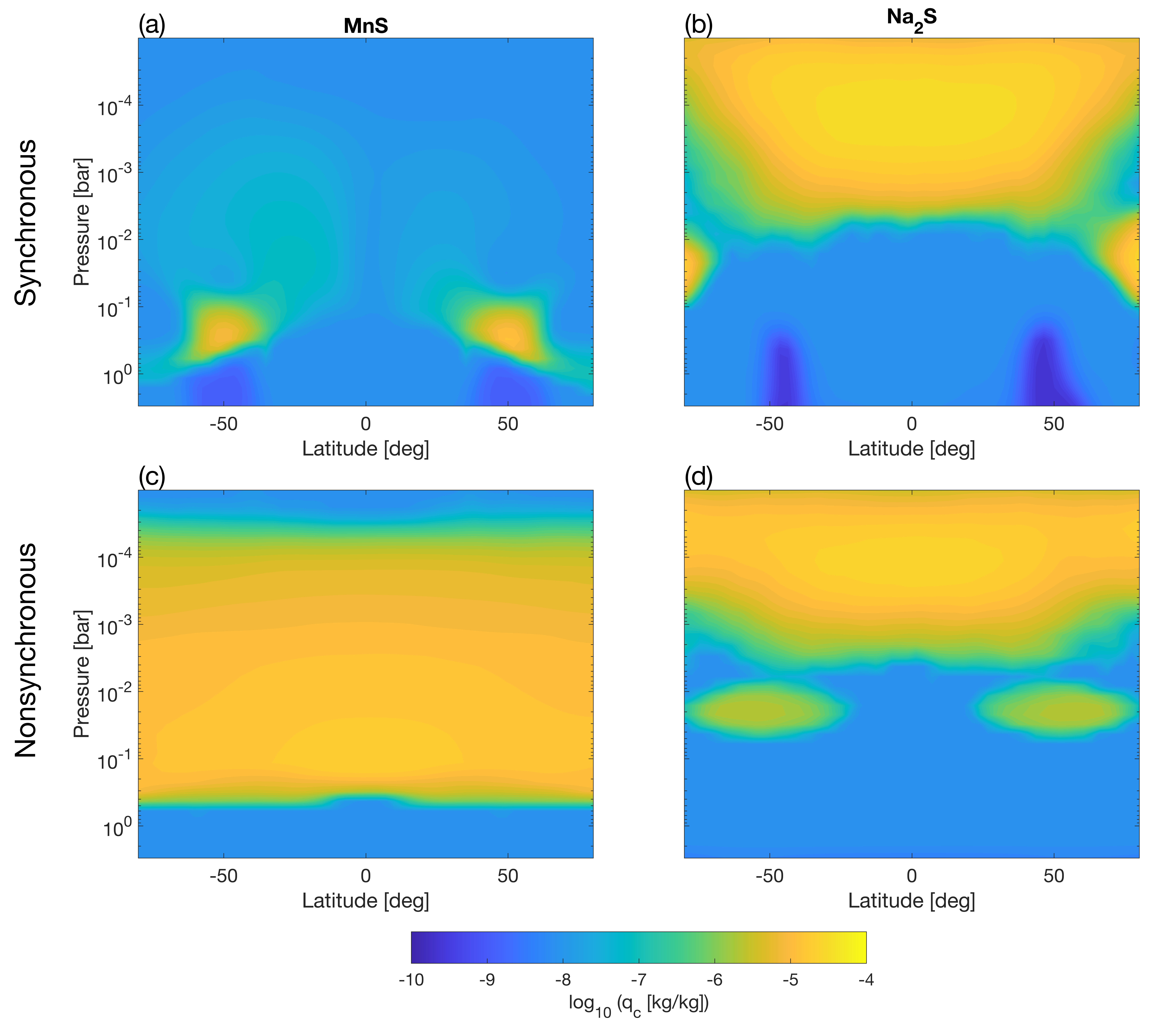}
    \caption{Zonal-mean cloud mass mixing ratio (log$_{10}$(q$_c$), kg\,kg$^{-1}$) of MnS (left) and Na$_2$S (right) versus latitude and pressure under synchronous (top panels) and nonsynchronous rotation (bottom panels). For the nonsynchronous case, one-orbit-mean fields during the eastward phase are shown.}
    \label{fig:cloud_compare}
\end{figure*}

\begin{figure*}[!t]
    \centering
    \includegraphics[width=0.8\textwidth]{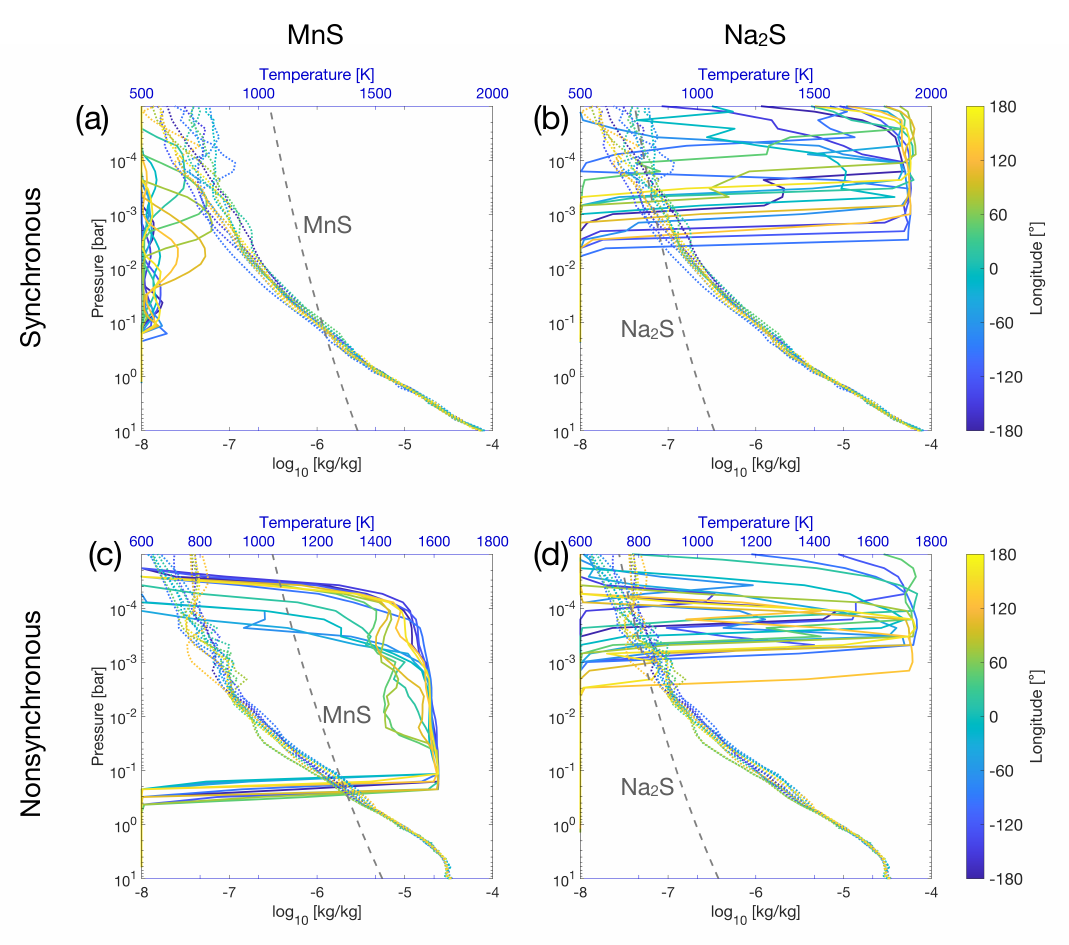}
    \caption{Condensation curves, temperature, and cloud mass mixing ratios for MnS (left) and Na$_2$S (right) under synchronous (top) and nonsynchronous (bottom) rotation. Dashed gray lines represent the condensation curves from \cite{morley2012} (K), dotted color lines show temperature profiles at the equator (K), and solid color lines indicate the cloud mass mixing ratio at the equator (log$_{10}(q_c$), kg\,kg$^{-1}$). Colors represent different longitudes.} 
    \label{fig:tp-cloud}
\end{figure*}

\begin{figure*}[!t]
    \centering
    \includegraphics[width=0.99\textwidth]{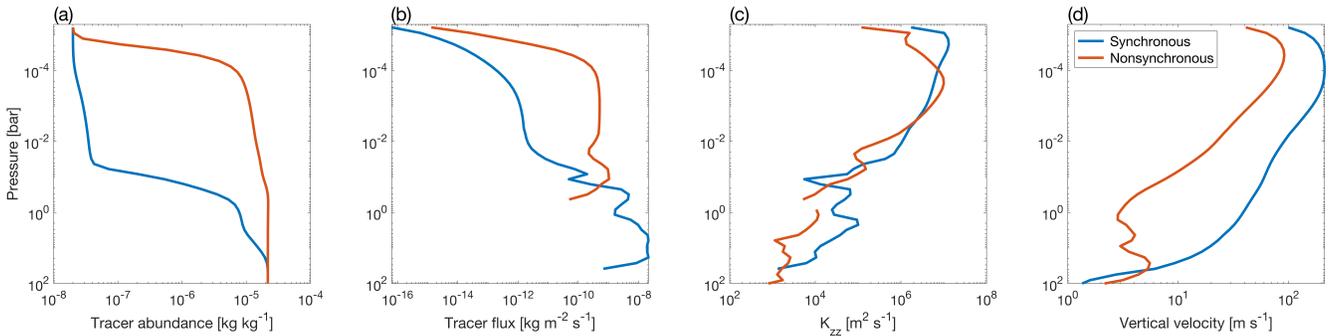}
    \caption{From left to right: global-mean MnS abundances (kg\,kg$^{-1}$), tracer flux of MnS (kg\,m$^{-2}$\,s$^{-1}$), GCM-derived vertical eddy diffusion coefficient $K_{zz}$ for MnS (m$^2$\,s$^{-1}$), and the root mean square of the vertical velocity (m\,s$^{-1}$) as functions of pressure. Blue and orange lines represent results for synchronous and nonsynchronous rotation, respectively. The profiles at pressures with negative tracer flux and $K_{zz}$ are truncated.}
    \label{fig:MnS-diag}
\end{figure*}

\subsubsection{Influence of clouds}\label{sec:nonsync_cloud}

Transmission spectra and phase curve observations suggest that clouds are ubiquitous in exoplanet atmospheres \citep[e.g.,][]{demory2013,kreidberg2014,helling2014,knutson2014,sing2016,benneke2019,malsky2021,feinstein2023}. Clouds can significantly affect radiative forcing through modulating atmospheric opacity, thereby shaping both the circulation and spectra properties of exoplanets \citep{marley2013,helling2019,parmentier2021,tan2021a,tan2021b}. 
In this study, we include two representative cloud species that are relevant at this effective temperature \citep{morley2012,parmentier2021}, MnS and Na$_2$S, to investigate their influence on the circulation and temperature patterns of high-obliquity mini-Neptunes. Details of cloud modeling are referred to Section \ref{sec:cloudscheme}.

Cloud distributions differ between the nonsynchronous and synchronous rotation cases, particularly for MnS clouds. Under synchronous rotation (Figure \ref{fig:cloud_compare}(a-b)), MnS clouds are limited and appear primarily in layers above the condensation level and below $p\sim 0.01$ bar. In contrast, Na$_2$S clouds are present globally above 10$^{-2}$ bar, with well-mixed clouds reaching low pressures to 10$^{-5}$ bar. Under nonsynchronous rotation (Figure \ref{fig:cloud_compare}(c-d)), however, MnS clouds are uniformly distributed across all latitudes and extend to layers within $10^{-4}\sim0.3$ bar. Meanwhile, Na$_2$S clouds are well-mixed to the upper atmosphere where $p<$10$^{-3}$ bar, which is similar to the distribution under synchronous rotation. 

In general, cloud coverage of the non-synchronously rotating case closely follows where the condensation curves permit, e.g., clouds predominantly form in regions where the local temperature falls below the condensation temperature of each species (Figure~\ref{fig:tp-cloud}). In the synchronously rotating case, Na$_2$S clouds also follow where the condensation curve permits, but the MnS cloud coverage is quite limited, even though they should be thermodynamically stable to form (the local temperatures are lower than the MnS condensation temperature, Figure~\ref{fig:tp-cloud}(a)). 
In an attempt to understand the MnS cloud mixing, we diagnose the isobaric tracer flux and estimate the vertical eddy diffusion coefficient $K_{zz}$ for MnS based on GCM outputs following \cite{parmentier2013,zhang2018mixing,komacek2019mixing,tan2022bd}. Assuming that the global-mean vertical transport of tracer can be approximated as a diffusion process, a global-mean diffusion coefficient $K_{zz}$ can be derived by relating the mean vertical tracer flux to the mean tracer vertical gradient:
\begin{equation}
    \langle \rho \omega q\rangle = K_{zz} \langle \rho \frac{\partial q}{\partial z} \rangle,
    \label{eq:kzz}
\end{equation}
where $z = -H\log (p/p_{\star})$ is the log-pressure coordinate, $p_{\star}$ is the reference pressure, $H$ is the scale height, $\omega$ is the vertical velocity in this coordinate, $\rho$ is gas density, $q=q_c + q_v$ is the mass mixing ratio of the tracer, and the brackets represent the horizontal averaging along isobars and time averaging. 

The limited MnS vertical extent under synchronous rotation is likely related to the poor large-scale transport. As shown in Figure \ref{fig:MnS-diag}, at pressures below $\sim$1 bar, the synchronous rotation case exhibits MnS tracer fluxes that are 1--3 orders of magnitude lower than those in the nonsynchronous case. The weaker large-scale mixing results in the reduced MnS abundances in the synchronous case.
Interestingly, the average vertical velocities above 10 bars in the synchronous case are several times stronger than in the non-synchronous case, in stark contrast to the weaker vertical mixing of the former. It suggests that the correlations between tracer fields and vertical velocity fields (the term $\langle \rho \omega q\rangle$ in Equation~(\ref{eq:kzz})) may matter more to the mixing than considering the magnitude of the vertical velocities alone \citep{holton1986}. The GCM-derived $K_{zz}$ profiles are remarkably similar between the two cases despite very different tracer fluxes and abundances. Of the two cases, the anti-correlation between the strength of the vertical velocities and the MnS cloud mixing, and the independence of $K_{zz}$ to MnS cloud mixing, indicate that large-scale mixing in close-in, slowly-rotating exoplanet atmospheres is more complex than what’s been suggested as a simple diffusive process. 
A more detailed investigation into how large-scale mixing influences tracer abundances in general is reserved for future work.

The temperature pattern in the presence of clouds is modulated by the interactions between cloud-radiative feedback and circulation. In the synchronous case, the influence of clouds is minimal (Figure \ref{fig:influence_10solar_cloud}(c) and the green line of Figure \ref{fig:nonsync_tempdiff}), primarily because Na$_2$S clouds, the dominant species in this case, mainly reflect shortwave radiation. However, the shortwave absorption mainly occurs between $10^{-2}$ and 3 bar with solar metallicity (Figure \ref{fig:flux_compare}), where Na$_2$S clouds are largely absent, thereby limiting their radiative effect. By contrast, under nonsynchronous rotation, clouds evidently warm the atmosphere above the photosphere, reducing the temperature contrast and weakening the zonal jets (Figure \ref{fig:influence_10solar_cloud}(d) and the brown line of Figure \ref{fig:nonsync_tempdiff}). This effect is mainly driven by dominant MnS clouds over the globe, which efficiently absorb and emit longwave radiation. 

With clouds included, QBO-like oscillations featuring alternating eastward and westward jets also develop under nonsynchronous rotation (Figure~\ref{fig:10solar_cloud_qbo}(b)). Throughout the oscillation cycle, MnS cloud coverage shows relatively small temporal variability, but Na$_2$S clouds exhibit evident variations, with higher mixing ratios in cooler phases and lower in warmer phases. This difference arises because Na$_2$S clouds form at lower pressures where temperature fluctuations are greater, making it more variable than MnS clouds (Figure~\ref{fig:tp-cloud}(d)).

\begin{figure*}
    \centering
    \includegraphics[width=0.95\textwidth]{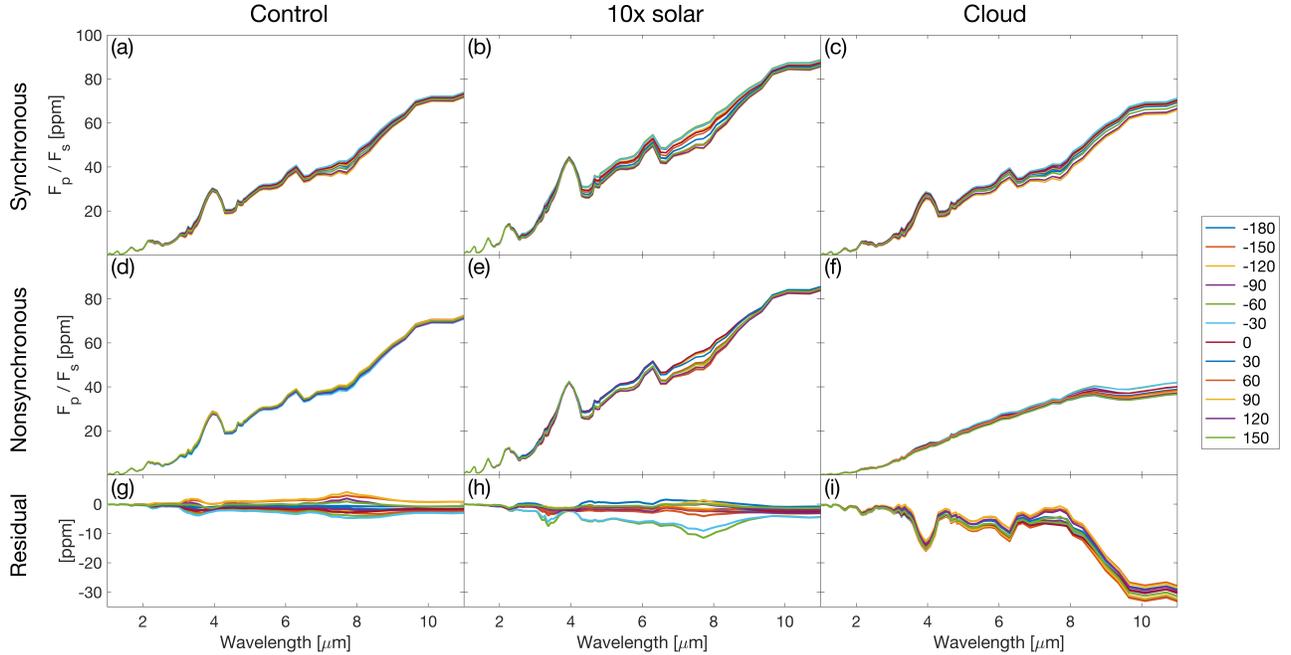}
    \caption{Thermal planet-to-star flux ratio as a function of wavelength from 1 to 11 $\mu$m for control cases (left), cases with 10$\times$ solar metallicity (middle), and cases including clouds (right). In each column, the top and middle panels show the flux ratios under synchronous and nonsynchronous rotation, while the bottom panel shows the residual difference between nonsynchronous and synchronous cases. In each panel, solid lines in different colors represent the flux ratios at phases ranging from -180$^{\circ}$ to 170$^{\circ}$ in 30$^{\circ}$ intervals. For the nonsynchronous cases, the flux ratios are shown at a tilting angle of 0$^{\circ}$ .}
    \label{fig:compare-fpfs-spectrum}
\end{figure*}

\section{Discussion}\label{sec:discuss}

\subsection{Observational signatures}\label{sec:spectra}
One of our main interests in this work is to examine the effects of different spin configurations on the atmospheric observations, similar to previous work \citep[e.g.,][]{rauscher2023}. To do so, we adopt the radiative transfer code PICASO to post-process observation signatures, including thermal emission and transmission spectra. 
Note that the sub-observed latitude varies with the tilting angle ($i$) between the spin axis and the line-of-sight under nonsynchronous rotation, which might impact the observational signatures. With an obliquity of $67^{\circ}$, the tilting angle and the sub-observed latitude range from $-67^{\circ}$ to $67^{\circ}$. Thus, we post-process observation signatures by assuming a group of tilting angles ($i$= 0$^{\circ}$, 20$^{\circ}$, 40$^{\circ}$, and 67$^{\circ}$) for each nonsynchronously rotating case.

\begin{figure*}
    \centering
    \includegraphics[width=0.75\textwidth]{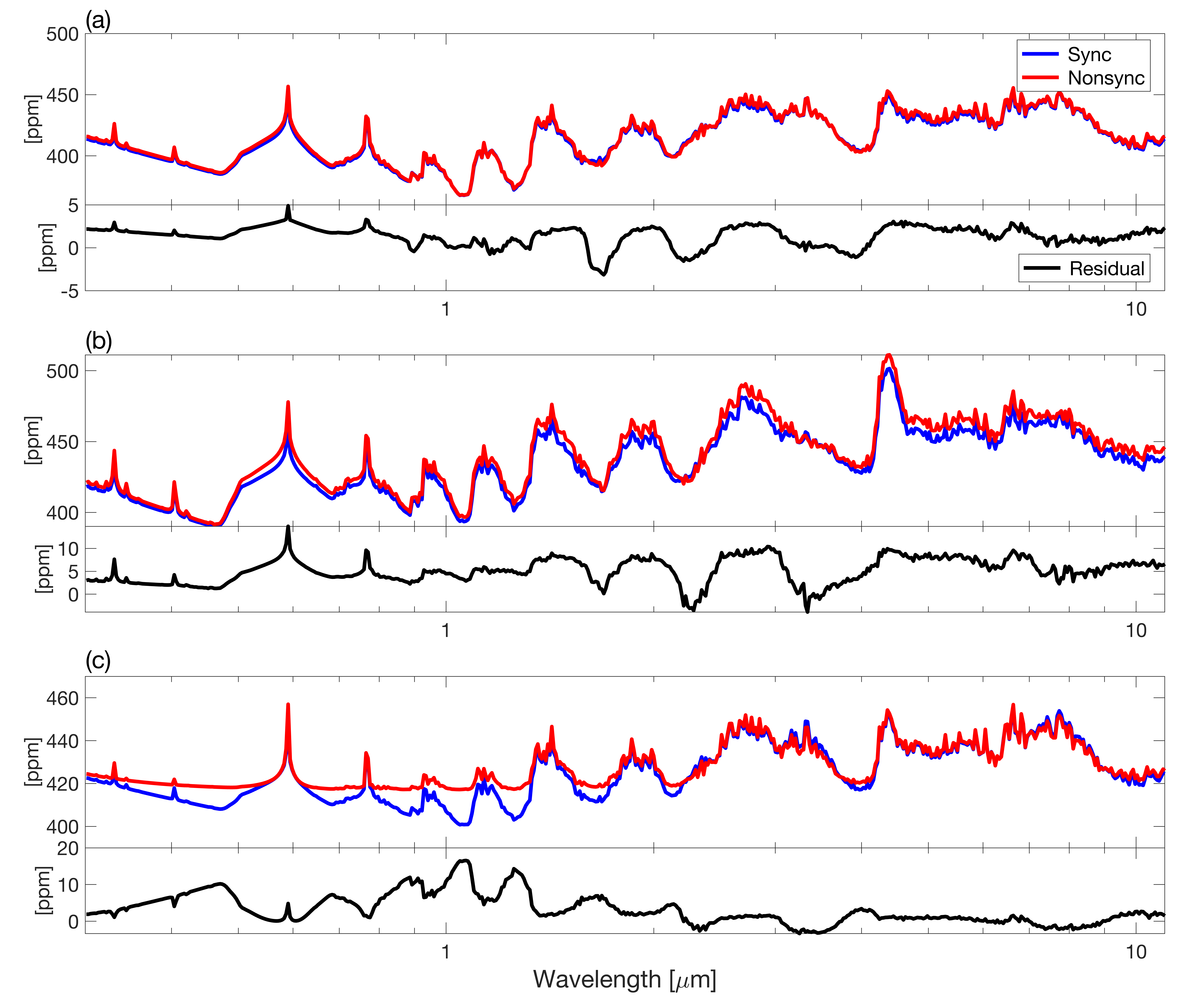}
    \caption{Transmission spectra for control cases (a), cases with 10$\times$ solar metallicity (b), and cases including clouds (c). In each panel, the upper subpanel shows the transit depth versus wavelength under synchronous (blue) and nonsynchronous rotation (red), while the lower subpanel shows the residual difference between the nonsynchronous and synchronous cases (black). For the nonsynchronous cases, the transmission spectra are shown at a tilting angle of 0$^{\circ}$.}
    \label{fig:compare-transit-spectrum}
\end{figure*}

Figure \ref{fig:compare-fpfs-spectrum} presents the thermal planet-to-star flux ratios for different cases and the residual difference between the nonsynchronous and synchronous cases within each group. Across the six scenarios, the flux ratios remain below 100 ppm, and the phase-to-phase differences are on the order of 10 ppm at most, regardless of tidal locking or not. Such limited phase variations result from the weak horizontal temperature gradients associated with the planet, which become even smaller under nonsynchronous rotation. 
Among these cases, those with 10$\times$ solar metallicity produce the largest flux ratios and phase variations (Figure \ref{fig:compare-fpfs-spectrum} (b \& e)), consistent with their larger horizontal temperature contrast near the photospheric layers (Figure \ref{fig:nonsync_tempdiff}). 
Clouds have only a minor effect on the spectra under synchronous rotation, as their influence on the temperature distribution is minimal and their photospheric coverage remains small (Figure \ref{fig:compare-fpfs-spectrum}(c)). In contrast, under nonsynchronous rotation, the flux ratios are reduced due to the dominant MnS cloud, which traps the thermal emission (Figure~\ref{fig:compare-fpfs-spectrum}(f)). 

For the cloudfree cases, the residual difference between nonsynchronous and synchronous cases remains below 10 ppm (Figure \ref{fig:compare-fpfs-spectrum}(g \& h)). When clouds are included, the residual difference increases slightly at specific wavelengths (e.g., $\sim$4 $\mu m$ and $\sim$10 $\mu m$) (Figure \ref{fig:compare-fpfs-spectrum}(i)), which is likely driven by strong scattering from cloud particles under nonsynchronous rotation \citep{wakeford2015,pinhas2017}.

The transmission spectra show only minor variations across all cases. Figure \ref{fig:compare-transit-spectrum} shows the wavelength-dependent transit depth for different cases, together with the residual difference between nonsynchronous  and synchronous cases within each group. For each group of simulations, the transit-depth difference remains at the level of $\sim$10 ppm, suggesting that the influence of planetary obliquity on transmission spectra is limited. Compared to the control group (Figure \ref{fig:compare-transit-spectrum}(a)), both the 10$\times$ solar metallicity and cloudy cases exhibit larger transit depths due to enhanced gas or cloud opacity (Figure \ref{fig:compare-transit-spectrum}(b \& c)). Under higher metallicity, the abundances of radiatively active gases rise proportionally, leading to overall higher opacity without altering the spectral features. In contrast, the transmission spectrum is slightly flattened in the cloudy cases, as the clouds introduce continuum opacity and then reduce the amplitudes of spectral features. The transit depth is slightly larger in the nonsynchronous cases, consistent with the presence of more abundant clouds (Figure \ref{fig:cloud_compare}).

It should be noted that for nonsynchronously rotating cases, the influence of various titling angles between the spin axis and the line-of-sight is minor on both thermal emission and transmission spectra, owing to the global WTG behavior of the planet.
Overall, the differences caused by high obliquity on both thermal emission and transmission spectra remain at the level of 10 ppm, which could be challenging to measure for current facilities, including JWST \citep{kempton2024}.

\subsection{Model limitations}\label{sec:limitation}
The atmospheric compositions of mini-Neptunes are expected to span a wide range \citep[e.g.,][]{valencia2013,zeng2019,misener2021,madhusudhan2020,madhusudhan2025}. These planets may host a rocky core surrounded by an H$_2$–He envelope (gas dwarfs) or contain substantial fractions of multicomponent, H$_2$O-rich ices and fluids (hycean worlds). In this study, we focus on the gas-dwarf scenario and model mini-Neptunes as planets enveloped by a thick H$_2$ atmosphere. If the atmosphere consisted of a multi-component mixture with higher metallicity, the corresponding increase in mean molecular weight would reduce the WTG parameter, which scales inversely with the square root of the mean molecular weight (see Equation (\ref{eq:WTGpara})). Nevertheless, even adopting a representative value of $\mu=10$, the resulting WTG parameter for those candidates in Table \ref{tab:list} remains much greater than unity, indicating that their atmospheric circulation resides in the WTG regime. This conclusion is consistent with \cite{innes2022}, which also predicts WTG behaviour for tidally locked, temperate, and slowly-rotating mini-Neptunes across different metallicity.

Photochemical haze, produced by UV-driven photochemical reactions at high altitudes, is a key opacity source in temperate atmospheres with equilibrium temperatures below 1000 K \citep{gao2020}. This makes it particularly relevant for most high-obliquity mini-Neptune candidates considered in this study (Table \ref{tab:list}). Such hazes can strongly influence the thermal structure, atmospheric circulation, and observational characteristics of these planets \citep[e.g.,][]{zhang2017b,horst2018}. GCM simulations of hot Jupiter show that the inclusion of photochemical hazes can generate additional dayside heating and temperature inversions at low pressures, thereby enhancing the day–night temperature contrast and increasing phase-curve amplitudes \citep{steinrueck2023}. For high-obliquity mini-Neptunes, which generally exhibit weak observable signals, incorporating photochemical haze in the models may therefore enhance the detectability of atmospheric features. 

\section{Conclusions}\label{sec:conclude}
In compact multiplanetary systems, secular spin-orbit resonances could induce high-obliquity, nonsynchronously rotating planets, which may strongly affect their atmospheric circulation. To explore the three-dimensional circulation and observational signatures of such planets, we focus on the representative case of K2-290 b, a potential high-obliquity (67$^{\circ}$), slow-rotating mini-Neptune, using ADAM GCM with post-processing by PICASO. Our results are expected to be applicable to a range of similar exoplanet systems listed in Table \ref{tab:list}, which share similar parameters and atmospheric dynamical regimes with K2-290 b.
Below, we summarize the key findings:

\begin{enumerate}
    
\item Regardless of synchronously rotating or not, the slow rotation rate, moderate equilibrium temperature and planetary radius of K2-290 b can result in a Weak-Temperature-Gradient (WTG) parameter much larger than unity ($\wedge \approx 14$), indicating a global WTG behavior and weak temperature gradients in the atmosphere \citep{pierrehumbert2019}, and our GCM simulations support this picture. 

\item Synchronous rotation produces broad, eastward superrotating jets at low and mid-latitudes (peak at $\sim$2800 m\,s$^{-1}$), which efficiently redistribute heat and yield weak day-night contrasts less than 100 K. The circulation with a nonsynchronous rotation exhibits a seasonal cycle due to high obliquity and the seasonal variation of stellar irradiation, as well as long-term variabilities with a characteristic period of $\sim$70 orbits. The long-term variability is caused by alternating eastward and westward jets with quasi-periodic temperature oscillations, resembling Earth’s quasi-biennial oscillation (QBO). The zonal jets reach 1200 m\,s$^{-1}$ near 0.01 bar during the eastward phase with a warmer equator, while in the westward phase, jets weaken to $-$200 m\,s$^{-1}$ and equatorial temperature drops below the poles. 

\item The QBO-like oscillations are driven by wave-mean flow interactions, where upward-propagating waves transport eddy momentum that accelerates zonal jets. In the eastward phase, maximum eddy-induced eastward acceleration occurs at the base of the eastward jet, producing its downward propagation, whereas the westward phase lacks this feature. These oscillations extend from low to mid latitudes due to K2-290 b’s slow rotation.

\item Higher atmospheric metallicity and clouds influence atmospheric circulation. Higher metallicity (10$\times$ solar) enhances gas opacity, producing a shallower photosphere. This strengthens radiative forcing, increases horizontal temperature contrasts, and drives stronger zonal jets. Clouds have minimal effect under synchronous rotation, because the dominant Na$_2$S clouds reside mostly outside the main shortwave absorption region. In contrast, under nonsynchronous rotation, MnS clouds warm the upper atmosphere, reduce temperature contrasts, and weaken zonal jets.

\item Both thermal emission and transmission spectra show weak observational signals, making detecting such planets challenging. The planet-to-star flux ratios remain below 100 ppm with phase variations less than 10 ppm. Similarly, the transit depths differ by less than 10 ppm between synchronous and nonsynchronous cases. Higher metallicity enhances both thermal emission and transit depths through increased photospheric temperature and gas opacity, while clouds suppress emission and flatten transmission spectra.

\end{enumerate}

\section*{Acknowledgments}
XT is supported by the National Natural Science Foundation of China (grant No. 42475131).  YL is supported by the National Natural Science Foundation of China (grant No. 42508019).
YS acknowledges support by the Lyman Spitzer Jr.\ Postdoctoral Fellowship at Princeton University and by the Natural Sciences and Engineering Research Council of Canada (NSERC) [funding reference CITA 490888-16].
This project is supported in part by Office of Science and Technology, Shanghai Municipal Government (grant Nos. 24DX1400100, ZJ2023-ZD-001). We acknowledge the computational support provided by the Siyuan-1 cluster, supported by the Center for High Performance Computing at Shanghai Jiao Tong University. We thank Dong Lai for the helpful discussion that initiated this work.

\appendix
\setcounter{equation}{0}
\renewcommand\theequation{\Alph{section}\arabic{equation}}
\renewcommand\thefigure{\Alph{section}\arabic{figure}}
\renewcommand\thetable{\Alph{section}\arabic{table}}
\setcounter{figure}{0}
\setcounter{table}{0}

\section{Supplementary simulations under zero and high Obliquity conditions}\label{ap:2cases}

\begin{figure*}
    \centering
    \includegraphics[width=0.7\textwidth]{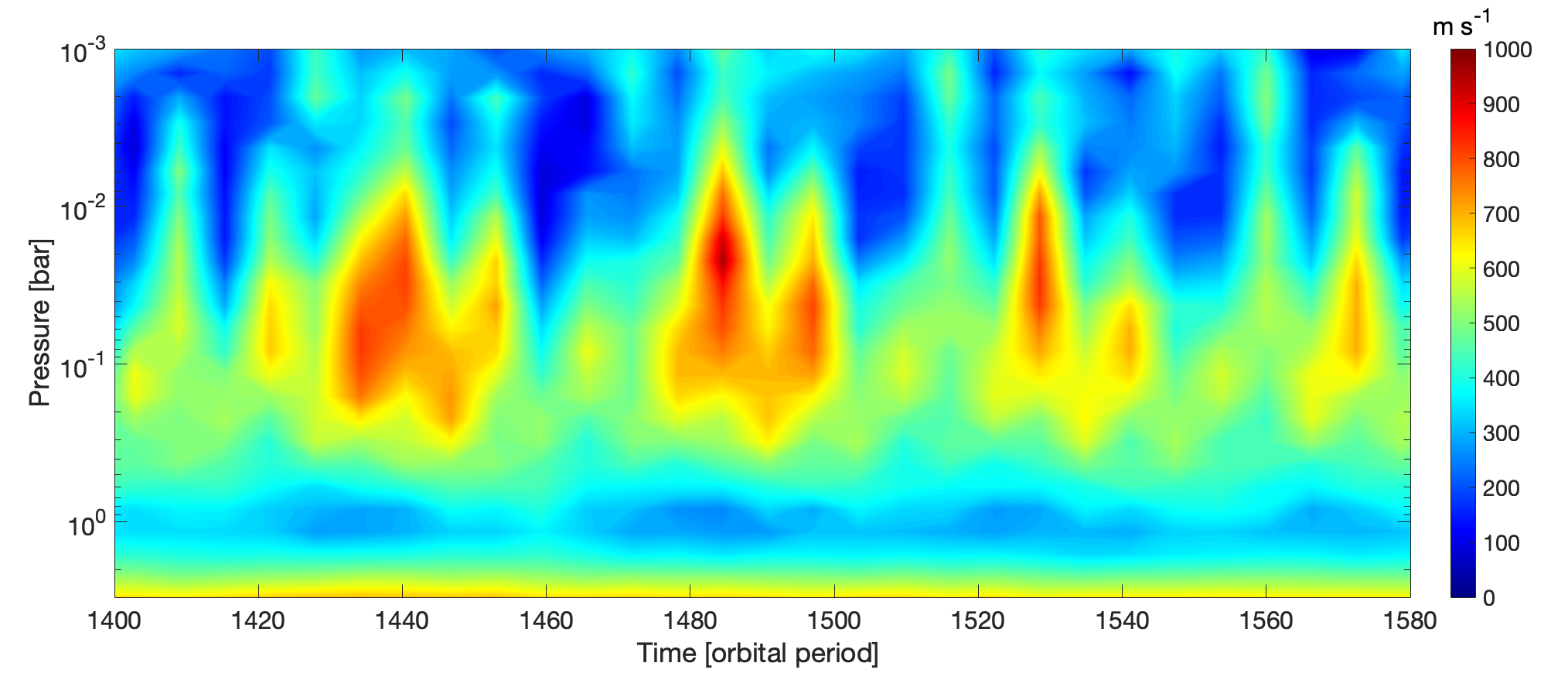}
    \caption{Timeseries of zonal-mean zonal wind versus pressure at 1.41$^{\circ}$N for the high-obliquity case ($\phi = 67^\circ$), with both the orbital and rotation periods set to 9.2 days.}
    \label{fig:sync_67obliq}
\end{figure*}

To qualitatively explore the roles of planetary obliquity and rotation in driving QBO-like oscillations, we briefly analyze simulation results for two cases: (1) a zero-obliquity case with rotation period $P_{\rm rot} = $ 13.4 days and orbital period $P_{\rm orb} = $ 9.2 days; (2) a high-obliquity ($67^{\circ}$) case with $P_{\rm rot} = P_{\rm orb} =$ 9.2 days. The first case draws on the work of \cite{penn2017} and \cite{penn2018}, who systematically investigated the atmospheric circulation of terrestrial planets across diverse orbital configurations under zero obliquity using shallower-water and idealized general circulation models, while the second case is based on an additional simulation conducted in this study (Figure \ref{fig:sync_67obliq}).

\cite{penn2017} and \cite{penn2018} demonstrated that in a moving-forcing system (i.e., where the substellar longitude changes with time), atmospheric zonal jets are sensitive to both planetary rotation and orbital period. These two parameters together determine the substellar velocity $s=a(\Gamma-\Omega)$, where $a$ is planetary radius, $\Gamma = 2\pi/P_{\rm orb}$ is the orbital rate, and $\Omega = 2\pi/P_{\rm rot}$ is the rotation rate. A positive $s$ indicates eastward drift of the substellar point, whereas a negative value corresponds to westward drift. The diurnal circle induced by the moving forcing excites internal waves, which deposit momentum into jets in the same direction as the moving substellar point. Specifically, eastward substellar velocities enhance atmospheric superrotation, while westward velocities weaken or even reverse this superrotation.

For our zero-obliquity case with $P_{\rm orb} = $9.2 days and $P_{\rm rot} = $ 13.4 days, the calculated substellar velocity is $s\sim50$ m\,s$^{-1}$, corresponding to eastward substellar drift. Under this velocity and rotation rate, persistent eastward zonal jets span from low to mid-latitudes (see Figure 4 \& Figure 6 in \cite{penn2018}) and no QBO-type oscillations are observed. Similarly, the high-obliquity case ($\phi=67^{\circ}$) with synchronous rotation shows no QBO-like behavior. As shown in Figure \ref{fig:sync_67obliq}, the atmosphere is dominated by permanent eastward jets, with no alternating eastward and westward winds.

In summary, QBO-like oscillations do not occur in either case (1) with only the asynchronous rotation or case (2) with only high obliquity. Instead, such oscillations appear in the nonsynchronous conditions described in the main text, suggesting that possibly only some parameter spaces are able to trigger appropriate upward-propagating waves that interact with the background mean flow to generate the QBO-like behaviors. 

\bibliography{main}{}
\bibliographystyle{aasjournal}

\end{document}